\documentclass[conference]{IEEEtran}
\usepackage{cite}
\usepackage{amsmath,amssymb,amsfonts}
\usepackage{algorithmicx}
\usepackage{graphicx}
\usepackage{textcomp}
\usepackage{xcolor}

\definecolor{added}{RGB}{0, 128, 0} % Green for additions
\definecolor{modified}{RGB}{255, 165, 0} % Orange for modifications

\usepackage[normalem]{ulem}
\usepackage{algpseudocode}
\usepackage{multicol}
\usepackage{multirow}
\usepackage{physics}
\usepackage{subcaption}
\usepackage[thmmarks]{ntheorem}
{
\theoremstyle{nonumberplain}
\theoremsymbol{\mbox{$\Box$}}
\newtheorem{proof}{Proof}
}
\newtheorem{definition}{Definition}
\newtheorem{theorem}{Theorem}%[section]
\newtheorem{proposition}{Proposition}%[section]
\newtheorem{lemma}{Lemma}%[section]
\newtheorem{example}{Example}%[section]
\newtheorem{corollary}{Corollary}%[section]

\def\BibTeX{{\rm B\kern-.05em{\sc i\kern-.025em b}\kern-.08em
    T\kern-.1667em\lower.7ex\hbox{E}\kern-.125emX}}
\begin{document}

\title{Verifying Randomized Consensus Protocols \\ with Common Coins
% {\footnotesize \textsuperscript{*}Note: Sub-titles are not captured in Xplore and
% should not be used}
% \thanks{Identify applicable funding agency here. If none, delete this.}
}

\author{\IEEEauthorblockN{Song Gao\IEEEauthorrefmark{4}\IEEEauthorrefmark{2}, Bohua Zhan\IEEEauthorrefmark{4}\IEEEauthorrefmark{2}, Zhilin Wu\IEEEauthorrefmark{4}\IEEEauthorrefmark{2}, 
Lijun Zhang\IEEEauthorrefmark{4}\IEEEauthorrefmark{2}\IEEEauthorrefmark{3}}
\IEEEauthorblockA{\IEEEauthorrefmark{4}\textit{University of Chinese Academy of Sciences, China} \\
\IEEEauthorrefmark{2}\textit{Institute of Software, Chinese Academy of Sciences, China}\\
\IEEEauthorrefmark{3}\textit{Institute of Intelligent Software, Guangzhou, China}\\
\{gaos, bzhan, wuzl,
zhanglj\}@ios.ac.cn
}
% \and
% \IEEEauthorblockN{Bohua Zhan}
% \IEEEauthorblockA{\textit{dept. name of organization (of Aff.)} \\
% \textit{name of organization (of Aff.)}}
% \and
% \IEEEauthorblockN{Zhilin Wu}
% \IEEEauthorblockA{\textit{dept. name of organization (of Aff.)} \\
% \textit{name of organization (of Aff.)}}
% \and
% \IEEEauthorblockN{Lijun Zhang}
% \IEEEauthorblockA{\textit{dept. name of organization (of Aff.)} \\
% \textit{name of organization (of Aff.)}}
% \and
% \IEEEauthorblockN{DOUBLE-BLIND}
}

\maketitle

\begin{abstract}
Randomized fault-tolerant consensus protocols with common coins are widely used in cloud computing and blockchain platforms. Due to their fundamental role, it is vital to guarantee their correctness. Threshold automata is a formal model designed for the verification of fault-tolerant consensus protocols. It has recently been extended to probabilistic threshold automata (PTAs) to verify randomized fault-tolerant consensus protocols. Nevertheless, PTA can only model randomized consensus protocols with local coins. 

In this work, we extend PTA to verify randomized fault-tolerant consensus protocols with common coins. Our main idea is to add a process to simulate the common coin (the so-called common-coin process). Although the addition of the common-coin process destroys the symmetry and poses technical challenges, we show how PTA can be adapted to overcome the challenges. We apply our approach to verify the agreement, validity and almost-sure termination properties of 8 randomized consensus protocols with common coins.
\end{abstract}


% {
% To-Do List:
% \begin{enumerate}
%     \item additional experiments on 'whether different system settings affect the verification speed.'
%     \item extendable to complex protocols? 
%     \item how to deal with time complexity
%     \item explain necessary TA concepts, and how a TA models the protocol and Byzantine processes
%     \item explain how the counterexample matches the attack, and how to identify automatically
%     \item intro, motivation and challenges
%     \item explain the formulae in Table III    
% \end{enumerate}
% }
\section{Introduction}\label{sec:intro}
Consensus is a fundamental problem in distributed computing, where a number of processes need to agree on some data values during computation. Consensus protocols are generally designed to be fault-tolerant or resilient, which means that they can withstand the existence of Byzantine or unreliable processes. Consensus protocols have many important applications in various fields, such as cloud computing and blockchains. As a result, there is a large body of work on designing and verifying consensus protocols~\cite{AVFC:Ben-Or:83,RBG:Rabin:83,VRBA:KwiatkowskaN:02,ASCPTA:KonnovLVW:17,VRCAURRA:BertrandKLW:21,HVBC:BertrandGKL:22,EASABA:AbrahamBY:22}. However, formal verification of consensus protocols remains a difficult challenge, especially when the protocols are probabilistic and/or make use of additional cryptographic primitives.

The correct consensus protocols must meet three conditions: agreement, validity and termination. The result of FLP impossibility~\cite{Impossible:FischerLP:85} states that there is no deterministic protocol that satisfies the consensus of an asynchronous distributed system, where any process can fail arbitrarily. Randomness provides a solution to reach consensus when the termination requirement is weakened to require termination with probability $1$. In this way, the FLP argument no longer prohibits consensus: non-terminating executions still exist, but collectively they can only occur with probability $0$.

Ben-or~\cite{AVFC:Ben-Or:83} and Rabin~\cite{RBG:Rabin:83} proposed the first randomized consensus protocols, which laid the foundation for subsequent designs and contributed significantly to the development of the field. Early versions of randomized consensus protocols make use of a \emph{local coin} for randomization, where each process throws the coin independently. As a consequence, they have an exponential expected number of rounds, making them of theoretical interest but of limited practical use. Rabin~\cite{RBG:Rabin:83} introduced an additional computational power called a \emph{common coin}, which delivers the same sequence of random bits $b_0, b_1, \ldots, b_r$ to all processes (each bit $b_i$ has the value $0$ or $1$ with probability $1/2$). The common coin is powerful as it can provide a constant expected number of rounds.

However, designing and proving the correctness of a randomized fault-tolerant distributed protocol is challenging. Additionally, there exist several attacks~\cite{ABABug:miller:18,DBLP:journals/corr/abs-1712-01367} against peer-reviewed %, award-winning, 
and even practically used 
 protocols. 
The attacks motivate formal verification of randomized fault-tolerant protocols. Threshold automata (TA)~\cite{DBLP:journals/iandc/KonnovVW17,ASCPTA:KonnovLVW:17} have been used extensively for the verification of fault-tolerant distributed protocols. The work of Bertrand et al.~\cite{VRCAURRA:BertrandKLW:21} proposes an extension of threshold automata by adding probability, probabilistic threshold automata (PTA). The consensus properties of protocols are reduced to queries on a one-round threshold automaton, which can be automatically checked using Byzantine Model Checker (ByMC)~\cite{BMC:KonnovW:18}. However, this approach assumes local coins (the randomness in each process is independent) and cannot be used directly for protocols involving common coins.

Extending PTAs to deal with common coins brings additional challenges. As processes are modeled as identical automata and the symmetry is essential in the TA and PTA theories, we cannot simply let one distinguished process toss the common coin. Moreover, throwing the common coin is assumed not to be subjected to Byzantine faults. Hence, a nontrivial extension of PTA is necessary for modeling common coins as well as extending {parameterized} verification methods to an asymmetric system. We propose such an extension, by adding an additional automaton for modeling the common coin, as well as extra shared variables for communication between the common coin automaton and other processes. We fully revisit the theories of threshold automata and probabilistic counter systems as well as adapt the theorems and proofs. We then reduce correctness and termination checks of the extended model to single-round queries on {non-probabilistic} threshold automata, which can be checked using ByMC. % An important subset of the queries correspond to the binding condition described above.

\paragraph{Computation Model}
In this work we consider asynchronous processes, which means that each process proceeds at its own pace, which may vary arbitrarily with time, and remains always unknown to the other processes. The system is made up of a finite set of $n$ asynchronous processes.%, namely $\{P_1, P_2, \ldots, P_i\}$. 

Processes communicate by exchanging messages through an asynchronous reliable point-to-point network. This network ensures that a message that has been sent is eventually received by its destination process without any loss, duplication or modification. Although there is no bound on the delay for message transfer, the network guarantees that the messages will be delivered correctly. The term ``point-to-point'' indicates that there is a bi-directional communication channel between each pair of processes, allowing a receiving process to identify the sender of a message. 
We assume that there are up to $t$ processes may exhibit Byzantine faults. %{(removed)}%A process that exhibits Byzantine behavior is called faulty; otherwise, it is correct. }
%A Byzantine process behaves arbitrarily: it may crash, fail to send or receive messages, send arbitrary messages, start in an arbitrary state, perform arbitrary state transitions, etc. 

This computation model is denoted $\mathcal{BAMP}_{n,t}[\emptyset]$ (BAMP stands for
Byzantine Asynchronous Message Passing). %As mentioned above, t
This model is both restricted with a resilience condition and enriched with additional computational power. More precisely, $\mathcal{BAMP}_{n,t}[n>3t, \textnormal{CC}]$ denotes that the computational model is enriched with a common coin with up to $t<n/3$ Byzantine processes in the system. 
% {(removed)}

% The model of a common coin has two properties: (1) $d$-\emph{Unpredictability}: the adversary cannot predict the coin value as long as at most $d$ parties started the protocol; (2) $\epsilon$-\emph{Good}:  for any value $v \in \{0,1\}$, all correct parties output $v$ with probability $\geq \epsilon$. 
% If a coin is $\frac{1}{2}$-good, we call it a \emph{strong coin}. In this paper, we consider only the protocols that employ strong coins.

$\epsilon$-\emph{Good} is an important property of the common coin abstraction: for any value $v \in \{0,1\}$, all
correct parties output $v$ with probability $\geq \epsilon$. If a coin is $\frac{1}{2}$-good, we call it a \emph{strong coin}. In this paper, we consider only the protocols that employ strong coins.

\paragraph{Contributions} The contributions of our work are as follows: {
\begin{itemize}
    \item We propose an extension of probabilistic threshold automata to incorporate common coins, and an extension of probabilistic counter system which removes the restrictions in the PTA method. 
    \item As the proposed extensions break the crucial symmetry in TA and PTA, we revisit several theorems for reducing the verification of multi-round, probabilistic specifications to the checks of single-round, non-probabilistic formulas, and re-prove their correctness in our extended model.
    \item We reduce the correctness and termination conditions for probabilistic threshold automata with a common coin to a set of queries for single-round threshold automata. The termination conditions include those for checking the \textit{binding} hyperproperty~\cite{EASABA:AbrahamBY:22}.
    \item Using the above framework and the ByMC tool, we verify a benchmark of 8 randomized distributed protocols using common coins. The verification is capable of reproducing the adaptive attack proposed in~\cite{ABABug:miller:18} against~\cite{SFABC:MostefaouiMR:14}, and verifying that the fixed versions are correct.
\end{itemize}
}
\paragraph{Outline of the paper} The remaining sections of the paper are as follows. Sect.~\ref{sec:motivation} gives a motivating example of common-coin-based protocol and its attack under an adaptive adversary. Sect.~\ref{sec:pta-cc-framework} introduces the framework of (probabilistic) threshold automata and extended counter systems with common coins. Sect.~\ref{sec:randomized-consensus} introduces randomized distributed consensus protocols and its correctness properties. Sect.~\ref{sec:verification-consensus} is the main part of the paper, describing our verification approach, reducing correctness of the protocol to proof obligations that can be checked using ByMC. We describe the experiments in Sect.~\ref{sec:experiments}, review related work in Sect.~\ref{sec:relatedwork} and conclude in Sect.~\ref{sec:conclusion} with a discussion of future work.

\section{Motivating Protocol and its Attack}\label{sec:motivation}
Common coin is a powerful abstraction to achieve randomized consensus, but the use of common coin in the presence of an \emph{adaptive} adversary makes it more difficult to reason about termination of the protocol. In particular, the work of Most\'efaoui et al.~\cite{SFABC:MostefaouiMR:14} proposed the first signature-free protocol, namely \texttt{MMR14}, to achieve asynchronous Byzantine consensus, with $O(n^2)$ messages and tolerance of $t<n/3$ Byzantine processes. However, an attack~\cite{ABABug:miller:18} was later found for the protocol. In order to demonstrate the abstraction of common coin and the structure of a multi-round BFT protocol, we describe the protocol and its attack in some detail.

The protocol \texttt{MMR14}  makes use of another abstraction called BV-broadcast, where each process broadcasts a binary value and obtains binary values in return. First, the $i$'th process $\mathcal{P}_i$ broadcasts its chosen value $v_i$. Then some value $v$ is received from $t+1$ processes and if $v$ is not broadcast, it again broadcasts $v$. Finally, when some value $v$ is received from $2t+1$ different processes, it adds the value $v$ to the set $\mathit{bin\_values}$ of values it received.

The complete consensus protocol proceeds in a number of rounds. Each process begins with a proposed value $v_i$, and let $\mathit{est_i}$ be the current estimate of the value to be decided upon, which is initialized to be $v_i$. In each round, each process performs BV-broadcast of $\mathit{est_i}$ using message of type \textsc{est}, and waits until the set $\mathit{bin\_values}$ becomes nonempty. Then it BV-broadcasts each value in $\mathit{bin\_values}$ using message of type \textsc{aux}. Next, it waits until receiving $n-t$ messages of type \textsc{aux}, carrying values in the set $\mathit{bin\_values}$, and then let $s$ be the value of the next throw of the common coin. Let $\mathit{values}$ be the set of values in these \textsc{aux} messages. If $\mathit{values}$ is a singleton $\{v\}$, then $v$ becomes the new $\mathit{est_i}$. Further, if $v=s$, then $v$ is decided upon. If $\mathit{values}$ contains both 0 and 1, then the new $\mathit{est_i}$ is the coin value $s$.

\begin{figure}[tbp]
    \centering
    \begin{algorithmic}[1]
        \State $\mathit{est}_i \gets \mathit{init}$;~$r_i \gets 0$;
        \While{true}
              \State $r_i \gets r_i+1$; ~ $\operatorname{BV\_broadcast}$(\textsc{est},$r_i$,$\mathit{est}_i$);
              \State \textbf{wait until} ($\mathit{bin\_values}_i[r_i] \neq \emptyset $);
              \State $\operatorname{broadcast}$(\textsc{aux},$r_i$,$w$), where $w \in \mathit{bin\_values}_i[r_i]$;
              \State \textbf{wait until} ($\exists$ a set of $(n-t)$ (\textsc{aux},$r_i$,$x$) messages from distinct processes such that $\mathit{values}_i \subseteq \mathit{bin\_values}_i[r_i] $, where $\mathit{values}_i$ is the set of values in the messages);
              \State $s \gets \operatorname{random}()$; \quad \% common coin \%
              \If{ ($\mathit{values}_i=\{v\}$) \quad \% i.e., $|\mathit{values}_i|=1$ \% \\ \qquad }{
                    $\mathit{est}_i \gets v$; 
                    \If{ $v=s$ }{ $\operatorname{decide}$($v$) if not yet done;} \EndIf
              }\Else{ $\mathit{est}_i \gets s$; }
              \EndIf
        \EndWhile

    \end{algorithmic}
    \caption{Randomized Consensus 
    Protocol \texttt{MMR14} for Correct Process $\mathcal{P}_i$}
    \label{fig:algo-mmr14}
\end{figure}

The paper~\cite{SFABC:MostefaouiMR:14} gives a proof of termination of the protocol in expected finite number of rounds. The basic idea is by dividing the protocol into two distinct phases. In the first phase, it is guaranteed that all correct processes will eventually have the same value. It follows from the observation that at the end of each round, a correct process updates its value to either the only majority $v$ (if it exists) or the common coin result. Then with probability $1/2$ the common coin result is equal to $v$ and all correct processes get the same value $v$ at the end of this round. Thus, the expected number of rounds for this to occur is bounded by $2$. Moving on to the second phase, all correct processes broadcast the agreed-upon value $v$ and we can easily conclude that it happens in every later round. Then with probability $1/2$ the common coin result aligns with $v$ and consequently all correct processes decide $v$ in this round. Similarly, the expected number of rounds for the second phase is bounded by $2$. Combining the two phases, the expected termination time is four rounds.

However, this proof neglects the ability of an adaptive adversary to obtain the value of the common coin and then manipulates the behavior of the Byzantine processes as well as the schedule for sending messages to make the protocol never terminate. In more detail, consider a smallest system consisting of 3 correct processes $A_1, A_2, B_1$ and a Byzantine process $P_{byz}$, and at a round $k$, $A_1, A_2$ propose estimate value $0$ while $B_3$ proposes $1$. The adversary can delay $A_2$'s reception of all messages unboundedly, while $A_1$ and $B_1$ proceed until they have the same set $\textit{values}=\textit{bin\_values}=\{0,1\}$.
Therefore they both enter Line 12 and can only set their $est$ value to be the coin result $s$. Later $P_{byz}$ can manipulate the order of $A_2$ receiving messages, make its $values$ set equal to $\{1-s\}$ and finally set its $est$ value as $1-s$. At the end of this round, two correct processes have their new $est$ value $s$ and one has $1-s$, which is {either the same or dual} to the initial state. It indicates that no correct process can ever terminate.

This attack was later fixed by the authors in the journal version of the paper~\cite{SFABBC:MostefaouiMR:15}. Moreover, the work by Abraham et al.~\cite{EASABA:AbrahamBY:22} proposes a general framework for building protocols tolerating crash or Byzantine failures, and the implementation in the paper can also be viewed as a fixed version of \texttt{MMR14}. A key contribution of~\cite{EASABA:AbrahamBY:22} is proposing the \emph{binding} condition, which summarizes the property the protocol must satisfy in order to prevent the above attack. Intuitively, the binding condition states that in any round of an execution, by the time the first correct process accesses the common coin, there is already a value $b\in\{0,1\}$ such that in any extension of the execution, no process may output $b$ in the same round. Since the adversary gains knowledge of the common coin only when the first correct process accesses it, the adversary can no longer always manipulate the Byzantine processes to output the opposite value. A formal definition of the binding condition, in the language of threshold automata defined in our work, will be given in Sect.~\ref{subsec:termin}.

\begin{figure}[tbh]
    \centering
    \begin{algorithmic}[1]
        \State \textbf{input} $b_i \in \{0,1\}$;
        \State $\operatorname{broadcast}$ message $b_i$;
        \State \textbf{wait until} ($\exists ~d_i \in \{0,1\}$ is received $\left\lceil\frac{n+1}{2}\right\rceil$ times );
        \State $\operatorname{decide}(d_i)$
    \end{algorithmic}
    \caption{Naive Voting Protocol for Correct Process $\mathcal{P}_i$}
    \label{fig:algo-voting}
\end{figure}

\begin{figure}[htbp]
    \centering
    \includegraphics[width=0.4\textwidth]{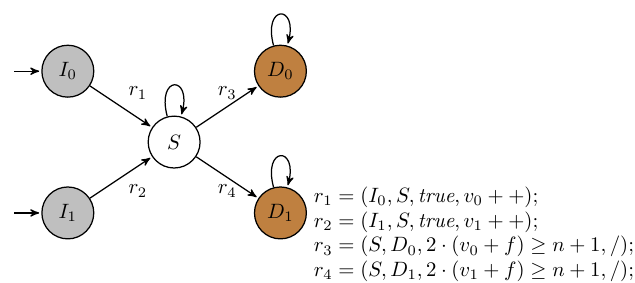}
    \caption{Threshold Automaton for Naive Voting}
    \label{fig:ta-voting}
\end{figure}

\section{The Framework of Probabilistic Threshold Automata Extended with Common Coins}
\label{sec:pta-cc-framework}

\subsection{Threshold Automata}
Let us start with threshold automata~\cite{ASCPTA:KonnovLVW:17,DBLP:journals/iandc/KonnovVW17,BMC:KonnovW:18}. A threshold automaton describes how a correct process runs in a concurrent 
system, and it usually contains local locations, variables and transition rules. Then the system can be abstracted as a counter system of multiple copies of such threshold automata, and its running states can be captured by the counters of locations.

\begin{example}
    Fig.~\ref{fig:algo-voting} shows a simple protocol of majority voting, and we model it with the threshold automaton shown in Fig.~\ref{fig:ta-voting}. We assume that there are $n$ processes in total, and $f$ is the number of Byzantine processes. We run $n-f$ instances of the threshold automaton; each instance is modeling a correct process.
    
    There are two initial locations $\{I_0, I_1\}$, which indicate the input of the process, and two final locations $\{D_0, D_1\}$ for its decision value. The shared variables $v_0, v_1$ represent the number of messages sent by the correct processes. The transition rules show steps of the protocol: $r_1, r_2$ for Line 2 and $r_3, r_4$ for Line 3. Let $c_0$ be the configuration where $v_0=v_1=0$, and all counters are $0$ expect the counter of $I_0$ equals $n-f$.
    % $\kappa[I_0]=n-f$. 
    This configuration corresponds to a concurrent system where all correct processes have the same input $0$ and have not yet broadcast any message.

    Byzantine processes are not directly modeled in the automaton. However, Byzantine behaviors can be captured by their impact on the transitions of correct processes and on the threshold guards. For instance, assume that $\mathcal{P}_1, \mathcal{P}_2, \mathcal{P}_3$ involve in the naive voting protocol and $\mathcal{P}_3$ is Byzantine. $\mathcal{P}_1$ and $\mathcal{P}_2$ propose different binary values, therefore their threshold automata start with different initial states and reach  location $S$. $\mathcal{P}_1$ and $\mathcal{P}_2$ cannot proceed if they receive no message from $\mathcal{P}_3$, and if any of them decides value $0$, i.e. its automaton reaches location $D_0$, we can infer that Byzantine process $\mathcal{P}_3$ sends a message with value $0$ to it. In the threshold automaton, it is represented as a non-deterministic choice of $r_3$ and $r_4$. 
\end{example}

\subsection{Probabilistic Threshold Automata Extended with Common Coins}

\begin{figure*}[tbp]
    \centering   
    \begin{minipage}[b]{0.4\linewidth}
        \centering
        \includegraphics[width=\linewidth]{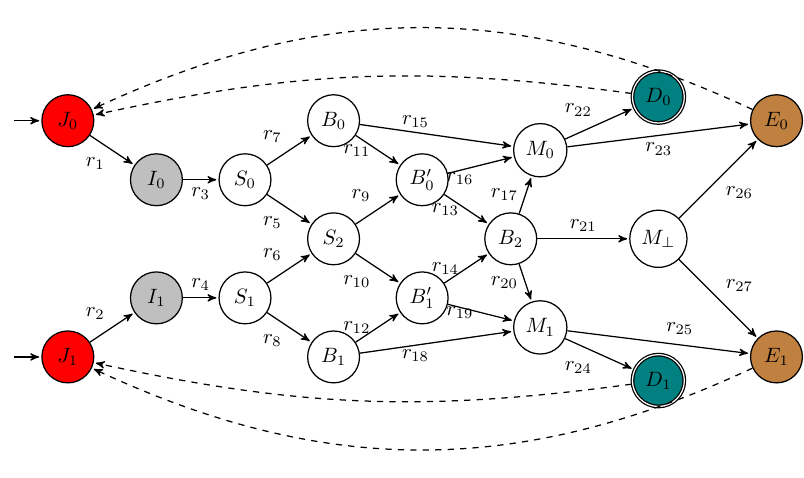}
        \caption*{(a)}
        \label{fig:np-ta}
    \end{minipage}
    \quad
    \begin{minipage}[b]{0.3\linewidth}
        \centering
        \includegraphics[width=\linewidth]{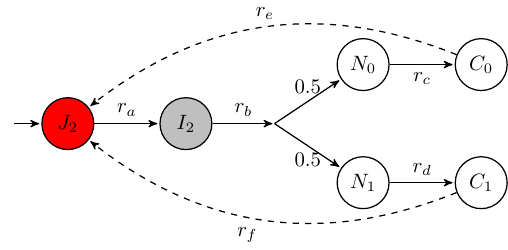}
        \caption*{(b)}
        \label{fig:cc-ta}
    \end{minipage}    

    \caption{Multi-round Threshold Automata for \texttt{MMR14}, with Self Loops Omitted}
    \label{fig:ta-MMR14}
\end{figure*}

We present the framework of probabilistic threshold automata and counter systems extended with common coins, illustrating the definitions in the example of Fig.~\ref{fig:ta-MMR14}, a model of \texttt{MMR14}. The automata rules are given in Table~\ref{tab:mmr14}.

\paragraph{Environments} (Probabilistic) threshold automata are defined relative to an environment $\textit{Env} = (\Pi, \textit{RC}, N)$, where $\Pi$ is a set of parameters that range over $\mathbb{N}_0$, \textit{RC} is the resilience condition, a formula in linear integer arithmetic over parameters. Intuitively, a valuation of $\Pi$ determines the number of different types of process in the system,  \textit{RC} defines the set of admissible parameters $\mathbf{P}_{\textit{RC}}= \{ \vb{p} \in \mathbb{N}_0^{|\Pi|}: \vb{p} \models \textit{RC} \}$. $N: \mathbf{P}_{\textit{RC}} \rightarrow \mathbb{N}^2_0$ is a function that maps a vector of admissible parameters to the number of modeled processes and common coins in the system. 

\begin{example}
    In the threshold automata of Fig.~\ref{fig:ta-MMR14}, the parameters are $n, f, t$ and {$cc$}, denoting the total number of  processes, the actual number of Byzantine processes, the maximum number of Byzantine processes while ensuring the correctness, and the number of common coins, respectively. The resilience condition is $n>5t ~\wedge~ t \geq f ~\wedge~ f \geq 0 ~\wedge~ {cc} \geq 1$, while the function $N$ is given by $(n,f,t, {cc} ) \mapsto (n-f, 1)$, as we model only $n-f$ correct processes and 1 common coin explicitly.
\end{example}

Next, we can define the (non-probabilistic) threshold automata for correct processes and the probabilistic threshold automata for common coins in the multi-round setting.

\paragraph{Threshold Automata for Correct Processes}
Formally, a (non-probabilistic) threshold automaton over an environment $(\Pi, \textit{RC}, N)$ is a tuple $\textbf{TA}^n = (\mathcal{L}^n, \mathcal{V}^n, \mathcal{R}^n)$, where

\begin{itemize}
    \item $\mathcal{L}^n$: a finite set of locations, which contains the following disjoint subsets:  initial locations $\mathcal{I}^n$, final locations $\mathcal{F}^n$, and border locations $\mathcal{B}^n$, with $|\mathcal{B}^n|=|\mathcal{I}^n|$;
    \item $\mathcal{V}^n$: a finite set of variables, including shared variables $\Gamma$ and coin variables $\Omega$; 
    \item $\mathcal{R}^n$: a finite set of rules; 
\end{itemize}

A simple guard is an expression of the form 
\[ b \cdot x \geq \Bar{a} \cdot \vb{p}^{\top} + a_0 \quad \textnormal{or} \quad b \cdot x < \Bar{a} \cdot \vb{p}^{\top} + a_0 ~, \]
where $x \in\Gamma$ is a shared variable, $\Bar{a} \in \mathbb{Z}^{|\Pi|}$ is a vector of integers, $a_0, b \in \mathbb{Z}$, and $\vb{p}$ is the vector of all parameters. Similarly, we can define a coin guard in the same form but on a coin variable.

A rule is a tuple $r=(\textit{from}, \textit{to}, \varphi, \vb{u})$, where $\textit{from, to} \in \mathcal{L}^n$ are the source and destination locations, $\varphi$ is a conjunction of guards, and $\vb{u} \in \mathbb{N}_0^{(|\Gamma|+|\Omega|)}$ is the update vector.

Note that for any rule $r=(\textit{from}, \textit{to}, \varphi, \vb{u})$, we have an additional restriction on $\varphi$ that it should be either a conjunction of simple guards or a conjunction of coin guards. We call a rule $r$ coin-based if $r.\varphi$ is a conjunction of coin guards; otherwise it is non-coin-based. Another restriction on $\vb{u}$ is necessary that the projection of $\vb{u}$ on the coin variables should be $\vb{0}$, that is, executing a rule in the threshold automata for correct processes should always keep the coin variables unchanged.

Threshold automata can model protocols with multiple rounds that follow the same code. Informally, a round starts from border locations and ends in final locations, and the code of a round is modeled by the transitions between initial locations and final locations. As $|\mathcal{B}^n|=|\mathcal{I}^n|$, we see that from each border location there is one rule towards an initial location, and it has the form $(\ell, \ell^\prime, \textit{true}, \vb{0})$ where $\ell \in \mathcal{B}^n$ and $\ell^\prime \in \mathcal{I}^n$. There are round-switch rules that let processes move from final locations of a certain round to border locations of the next round. They can be described as rules $(\ell, \ell^\prime, \textit{true}, \vb{0})$ where $\ell \in \mathcal{F}^n$ and $\ell^\prime \in \mathcal{B}^n$. The set of round-switch rules is denoted by $\mathcal{S}^n \subseteq \mathcal{R}^n$. A location belongs to $\mathcal{B}^n$ if and only if all incoming edges are in $\mathcal{S}^n$. Similarly, a location is in $\mathcal{F}^n$ if and only if there is only one outgoing edge and it is in $\mathcal{S}^n$.

A threshold automaton is called canonical if every rule $r$ that lies on a cycle ensures that $r.\vb{u}=\vb{0}$, and we consider only canonical ones in this work.

For binary consensus, every correct process, say $\mathcal{P}_i$, has an initial value $ \textit{init}_i \in \{0,1\}$, and its valid decision value (if any) should also be binary. Therefore, we can partition every set of locations $\mathcal{I}^n$, $\mathcal{F}^n$ and $\mathcal{B}^n$ into two subsets $ \mathcal{I}_{0}^n \uplus \mathcal{I}_{1}^n$, $\mathcal{F}_{0}^n \uplus \mathcal{F}_{1}^n$ and $ \mathcal{B}_{0}^n \uplus \mathcal{B}_{1}^n$,  respectively. For every $v \in \{0,1\}$, the partitions follow the 2 rules below:
\begin{enumerate}
    \item The processes that are initially in a location $\ell \in \mathcal{I}_v^n$ have the initial value $v$.
    \item Rules connecting locations from $\mathcal{B}^n$ and $\mathcal{I}^n$ respect the partitioning, i.e., they connect $\mathcal{B}_v^n$
and $\mathcal{I}_v^n$. Similarly, rules that connect the locations of $\mathcal{F}^n$ and $\mathcal{B}^n$ respect the partitioning.
\end{enumerate}
For common-coin-based protocols with a step for deciding a binary value, we can introduce two subsets, decision locations $\mathcal{D}_v^n \subseteq \mathcal{F}_v^n$, $v \in \{0,1\}$. Intuitively, a process is in $\mathcal{D}_v^n$ locations in round $k$ if and only if it decides $v$ in that round. Decision locations are accepting locations in a threshold automaton.

\begin{example}
    Fig.~\ref{fig:ta-MMR14}(a) depicts a threshold automaton with border locations $\mathcal{B}^n = \{J_0, J_1\}$, initial locations $\mathcal{I}^n = \{I_0, I_1\}$, final locations  $\mathcal{F}^n = \{E_0, E_1, D_0, D_1\}$, and decision locations $\mathcal{D}^n = \{D_0, D_1\}$. As $a_0, a_1, b_0, b_1$ are shared variables and $cc_0, cc_1$ are coin variables , there are 6 coin-based rules $r_{22},r_{23},r_{24},r_{25},r_{26}$ and $r_{27}$. The round-switch rules are represented by dashed arrows, and the self loops are omitted.
\end{example}

\paragraph{Probabilistic Threshold Automata for Common Coins}
A probabilistic threshold automaton for a common coin over an environment $(\Pi, \textit{RC}, N)$ is $\mathbf{PTA}^c = (\mathcal{L}^c, \mathcal{V}^c, \mathcal{R}^c)$, which extends the definition of threshold automata in the part of rules. 

Here a rule is a tuple $r=(\textit{from}, \delta_{to}, \varphi, \vb{u})$, where $\textit{from} \in \mathcal{L}^c$ is the source location,  $\delta_{to} \in \mathtt{Dist}(\mathcal{L}^c)$ is a probabilistic distribution over the destination locations, $\varphi$ is a conjunction of simple guards, and {$\vb{u} \in \mathbb{N}_0^{(|\Gamma|+|\Omega|)}$} is the update vector. If there exists $\ell \in \mathcal{L}^c$ such that $r.\delta_{to}(\ell)=1$, we call the distribution a \textit{Dirac} distribution and the rule a \textit{Dirac} rule $r=(\textit{from}, \ell, \varphi, \vb{u})$. 

Note that we remove the restriction on the occurrence of non-\textit{Dirac} rules, while~\cite{VRCAURRA:BertrandKLW:21} requires that the destination locations of all non-\textit{Dirac} rules should be in $\mathcal{F}^c$. 
The restrictions on rules are different in probabilistic threshold automata for common coins: $r.\varphi$ should only involve simple guards, and $r.\vb{u}$ cannot modify the values of shared variables. 

\begin{example}
    Fig.~\ref{fig:ta-MMR14}(b) shows a probabilistic threshold automaton for the common coin in \texttt{MMR14}. We have border locations $\mathcal{B}^c = \{J_2\}$, initial locations $\mathcal{I}^c = \{I_2\}$, and final locations  $\mathcal{F}^c = \{C_0, C_1\}$. The only non-$Dirac$ rule is $r_{b}$.% and it does not lie at the end of a round.
\end{example}

Finally, given $\textit{Env} = (\Pi, \textit{RC}, N)$, the non-probabilistic threshold automaton $\textbf{TA}^n = (\mathcal{L}^n, \mathcal{V}^n, \mathcal{R}^n)$ for correct processes and the probabilistic threshold automaton $\textbf{PTA}^c = (\mathcal{L}^c, \mathcal{V}^c, \mathcal{R}^c)$ for common coins, we have that they share the same set of variables, i.e., $\mathcal{V}^n = \mathcal{V}^c$, and their sets of locations and rules are naturally disjoint. For simplicity, we write  $\mathcal{L}$ for $ \mathcal{L}^n \cup \mathcal{L}^c$, $\mathcal{V} $ for $ \mathcal{V}^n$  and $\mathcal{R} $ for $ \mathcal{R}^n \cup \mathcal{R}^c$. Note that $(\mathcal{L,V,R})$ does not form a probabilistic threshold automaton for this system.

\begin{table}[btp]
	\centering
	\caption{The Rules of the Multi-round Threshold Automaton for \texttt{MMR14}}
	\begin{tabular}{l c c}
		\hline\hline\noalign{\smallskip}	
		Rules & Guard & Update  \\
		\noalign{\smallskip}\hline\noalign{\smallskip}
		  $r_1$, $r_2$ & \textit{true} & - \\
		  $r_3$ & \textit{true} & $b_0$\small$++$ \\
		  $r_4$ & \textit{true} & $b_1$\small$++$ \\
            $r_6$,$r_{12}$ & $b_0 \geq t+1-f$ & $b_0$\small$++$\\ 
            $r_5$,$r_{11}$ & $b_1 \geq t+1-f$ & $b_1$\small$++$\\ 
		  $r_7$, $r_9$ & $b_0 \geq 2t+1-f$ & $a_0$\small$++$ \\
		  $r_8$, $r_{10}$ & $b_1 \geq 2t+1-f$ & $a_1$\small$++$ \\
    {$r_{13}$} & $b_1 \geq 2t+1-f$ & - \\
		  {$r_{14}$} & $b_0 \geq 2t+1-f$ & - \\
		  $r_{15}$, $r_{16}$, $r_{17}$ & $a_0 \geq n-t-f$ & - \\
		  $r_{18}$, $r_{19}$, $r_{20}$ & $a_1 \geq n-t-f$ & - \\

		  $r_{21}$ & $a_0 + a_1 \geq n-t-f \wedge a_0 \ge 0 \wedge a_1 \ge 0$ & - \\
    
		  $r_{22}$, $r_{25}$, $r_{26}$ & $cc_0 > 0 $ & - \\
            $r_{23}$, $r_{24}$, $r_{27}$ & $cc_1 > 0 $ & - \\\hline
            
		  $r_{a}$, $r_{b}$ & \textit{true} & - \\
            $r_{c}$ & \textit{true} & $cc_0$\small$++$ \\
            $r_{d}$ & \textit{true} & $cc_1$\small$++$ \\
            % self loop & \textit{true} & - \\
		\noalign{\smallskip}\hline
	\end{tabular}
	\label{tab:mmr14}  
\end{table}

\subsection{Extended Probabilistic Counter Systems}\label{sec:prob-countersys}
The semantics of the probabilistic multi-round system is an infinite-state Markov decision process (MDP). Given an environment $\textit{Env} = (\Pi, \textit{RC}, N)$, a  threshold automaton $\textbf{TA}^n = (\mathcal{L}^n, \mathcal{V}^n, \mathcal{R}^n)$ for correct processes and a probabilistic threshold automaton $\textbf{PTA}^c = (\mathcal{L}^c, \mathcal{V}^c, \mathcal{R}^c)$ for the common coin, we define the semantics, called counter system $\textit{Sys}(\textbf{TA}^n, \textbf{PTA}^c)$ over \textit{Env}, to be the infinite-state
MDP $(\Sigma, I, \textbf{Act}, \Delta)$, where

\begin{itemize}
    \item $\Sigma$: the set of configurations,
    \item $I \subseteq \Sigma$ : the set of initial configurations,
    \item $\textbf{Act} = \mathcal{R} \times \mathbb{N}_0$: the set of actions labelled by round numbers,
    \item $\Delta: ~\Sigma \times \textbf{Act} \rightarrow \texttt{Dist}(\Sigma)$ is the probabilistic transition function.
\end{itemize}

\paragraph{Configurations} A configuration is a tuple $c=(\mathbf{\kappa, g, p})$, where the function $c.\mathbf{\kappa}: ~\mathcal{L} \times \mathbb{N}_0 \to \mathbb{N}_0$ describes the values of the location counters in each round, $c.\mathbf{g}: ~\mathcal{V} \times \mathbb{N}_0 \to \mathbb{N}_0$ defines the values of the variables in each round, and the vector $c.\mathbf{p}$ shows the values of the parameters. We denote the vector $(\mathbf{g}[x,k])_{x \in \mathcal{V}}$ of all variables in round $k$ by $\mathbf{g}[k]$, and denote the vector $(\kappa[\ell,k])_{\ell \in \mathcal{L}}$ of location counters in round $k$ by $\kappa[k]$.

A configuration is initial if all processes as well as the common coin are in initial locations of round $0$, and all variables evaluate to $0$. 
A threshold guard evaluates to true in a configuration $c$ for a round $k$, written $c,k \models \varphi$, if for all its conjuncts $b \cdot x \geq \Bar{a} \cdot \mathbf{p}^\top +a_0$, it holds that $ b \cdot c.\mathbf{g}[x,k] \geq \Bar{a} \cdot (c.\mathbf{p}^\top) +a_0 $, and similarly for conjuncts of the other form.

\paragraph{Actions and Probabilistic Transition Function} An action $\alpha=(r,k) \in \mathbf{Act}$ stands for the execution of a rule $r \in \mathcal{L}$ in round $k$ by a single process. 
% We write $\alpha.\textit{from}$ for $\alpha.r.\textit{from}$, $\alpha.\sigma_{to}$ for $\alpha.r.\sigma_{to}$, etc.

We say an action $\alpha$ unlocked in a configuration $c$ if the guard of its rule evaluates to true in round $k$, that is, $c, k \models r.\varphi$. An action $\alpha=(r,k)$ is applicable to a configuration $c$ if $\alpha$ is unlocked in $c$ and the location counter value at round $k$ for the source location of its rule is at least $1$, formally, $c.\kappa[r.\textit{from},k] \geq 1$. When an action $\alpha$ is applicable to configuration $c$ and $\ell$ is a potential destination location for the probabilistic action $\alpha$, we denote the resulting configuration as $\textit{apply}(\alpha, c, \ell)$. In this resulting configuration, the parameters remain unchanged, the variables are updated according to the update vector $\alpha.\vb{u}$, and  the values of location counters are modified in a natural way: in round $\alpha.k$, the counter of source location $\alpha.\textit{from}$ decreases by $1$, the counter of destination location $\ell$ increases by $1$, and the others remain unchanged.

The probabilistic transition function $\Delta$ is defined such that for any two configurations $c$ and $c^\prime$, and for any action $\alpha$ applicable to $c$, we have 
\[\Delta(c, \alpha)\left(c^{\prime}\right)=\left\{\begin{array}{ll}
\alpha.\sigma_{to}(\ell) & \qquad\text { if } \textit{apply}(c,\alpha,\ell)=c^{\prime} \\
0 & \qquad\text{ otherwise }
\end{array}\right.\]

\subsection{Non-probabilistic Extended Counter Systems}

Given a probabilistic threshold automaton \textbf{PTA} over an environment, we can replace probability with non-determinism and get a non-probabilistic threshold automaton \textbf{TA}$_{\textnormal{PTA}}$.
\begin{definition}
    Given a $\mathbf{PTA} = (\mathcal{L,V,R})$ over an environment \textit{Env}, its non-probabilistic threshold automaton is $\mathbf{TA}_{\textnormal{PTA}} = (\mathcal{L,V}, \mathcal{R}_{np})$ over \textit{Env} where the set of rules $\mathcal{R}_{np} $ is defined as $ \{ r_{\ell} = (\textit{from}, \ell, \varphi, \vb{u}) ~|~ r = (\textit{from}, \delta_{to}, \varphi, \vb{u}) \in \mathcal{R} \wedge \ell \in \mathcal{L} \wedge \delta_{to}(\ell) >0 \}$. 
\end{definition}
In words, $\mathcal{R}_{np}$ contains all \textit{Dirac} rules in $\mathcal{R}$, and turns all probabilistic branches of non-\textit{Dirac} rules into non-deterministic \textit{Dirac} rules. 
We write \textbf{TA} for \textbf{TA}$_{\textnormal{PTA}}$ when the automaton \textbf{PTA} is clear from the context, e.g., the non-probabilistic threshold automaton of \textbf{PTA}$^c$ is \textbf{TA}$^c$.

Given an environment $\textit{Env} $~=~$ (\Pi, \textit{RC}, N)$, a  threshold automaton $\textbf{TA}^n$ for correct processes and a probabilistic threshold automaton $\textbf{PTA}^c$ for the common coin, we can first get the non-probabilistic \textbf{TA}$^c$, and then define an infinite non-probabilistic counter system $\textit{Sys}_\infty(\textbf{TA}^n, \textbf{TA}^c)$ over \textit{Env}, to be the tuple $(\Sigma, I, \textbf{Act}^\prime, R)$. The set of configurations and initial configurations $\Sigma, I$ are defined as in Sect.~\ref{sec:prob-countersys}. An action $t \in \textbf{Act}^\prime$ is a tuple $(r,k) \in (\mathcal{R}^n \cup \mathcal{R}^c_{np})\times \mathbb{N}_0$, and $R$ is the transition relation. Two configurations $c_0,c_1$ are in the transition relation, i.e., $(c_0,c_1) \in R$ if and only if there exists an action $t$ such that $t(c_0)=c_1$.

In the non-probabilistic counter system, a (finite or infinite) sequence of transitions is called schedule, and it is often denoted by $\tau$. A schedule $\tau = t_1, t_2, \ldots, t_{|\tau|}$ is applicable to a configuration $c$ if there is a sequence of configurations $c_0, c_1, \ldots, c_{|\tau|}$ such that for every $1 \leq i \leq |\tau|$ we have that $t_i$ is applicable to $c_{i-1}$ and $c_i = t_i(c_{i-1})$. Given a configuration $c_0$ and a schedule $\tau$, we denote by $\textit{path}(c_0, \tau)$ a path $c_0, t_1, c_1, \ldots, t_{|\tau|}, c_{|\tau|}$ where $t_t(c_{i-1}) = c_i$ for every $1 \leq i \leq |\tau|$. Similarly we define an infinite schedule $\tau$ and an infinite path also denoted by $\textit{path}(c_o, \tau)$. 

% Observe that since every transition in $\textit{Sys}_{\infty}(\mathbf{TA}^n , \mathbf{TA}^c)$ comes from an action in $\textit{Sys}(\mathbf{TA}^n, \mathbf{PTA}^c)$, every path in $\textit{Sys}_{\infty}(\mathbf{TA}^n , \mathbf{TA}^c)$ corresponds to a valid path in $\textit{Sys}(\mathbf{TA}^n, \mathbf{PTA}^c)$.

An infinite path is fair if no transition is applicable forever from some point on. Equivalently, when a transition is applicable, eventually either its guard becomes false, or all processes leave its source location.

\subsection{Adversaries}
The non-determinism is usually resolved by a so-called adversary. We denote by $\Sigma^+$ the set of all non-empty sequences of configurations. {An adversary is a function $a: \Sigma^{+} \rightarrow$ \textbf{Act}, which given a sequence of configurations $\xi \in \Sigma^+$ selects an action applicable to the last configuration of $\xi$.} Given a configuration $c$ and an adversary $a$, we generate a family of paths, depending on the outcomes of non-Dirac transitions, and we denote this set by $paths(c, a)$. An adversary $a$ is fair if all paths in $paths(c, a)$ are fair.

The Markov Decision Process (MDP) $\textit{Sys}(\textbf{TA}^n, \textbf{PTA}^c)$ over \textit{Env}  together with an initial
configuration $c$ and an adversary $a$ induce a Markov chain, written as $\mathcal{M}^c_a$. We denote by $\mathbb{P}^c_a$ the probability measure over infinite paths starting at $c$ in the latter Markov chain.

An adversary $a$ is round-rigid if it is fair, and if every sequence of actions it produces can be decomposed into a concatenation of sequences in the form $s_0 \cdot s_1 \cdot s_2 \ldots$, where the sequence $s_k$ contains only actions of round $k$. We denote the set of all round-rigid adversaries by $\mathcal{A}^R$.

The atomic propositions $\textnormal{AP}_k$ and stutter equivalence discussed in this work follow the definitions in~\cite{VRCAURRA:BertrandKLW:21}.

\section{Randomized Distributed Consensus protocols}
\label{sec:randomized-consensus}

The consensus problem was first introduced by
Lamport et al.~\cite{DBLP:journals/toplas/LamportSP82}. It can be stated in a basic, generic manner: One or more processes may propose some value. How do we get a collection of correct processes to agree on exactly one of those proposed values?

In binary cases, assuming each correct process $p_i$ proposes a value $v_i \in \{0,1\}$, each of them has to decide a binary value.
\begin{definition}[Consensus]
     Consensus is reached if the following properties hold:
    \begin{itemize}
        \item Agreement: No two correct processes decide different values.
        \item Validity: A decided value was proposed by a correct process.
        \item Termination: Each correct process decides.
    \end{itemize}
\end{definition}

The famous FLP impossibility shows that no deterministic consensus protocol can be possible in asynchronous settings as soon as one node may crash. Ben-or~\cite{AVFC:Ben-Or:83} and Rabin~\cite{RBG:Rabin:83} are the first to show that the impossibility can be circumvented via randomness. In this paper, we focus on randomized binary consensus protocols in asynchronous systems, where randomization is provided by common coins. \textit{Randomized Consensus} is then defined by \emph{Agreement, Validity}, plus the following \emph{Almost-sure Termination} property: Each correct process decides with probability 1. For round-based protocols, consider the event $\mathcal{E}(i,r)$: process $p_i$ decides by round $r$. Then this termination property is re-stated as:
    for any correct process $p_i$, we have 
    $\textstyle \lim_{r \to +\infty} \mathbb{P}(\mathcal{E}(i,r)) = 1 $.

 Now we can express the specifications in \textnormal{LTL}$_{-\textnormal{X}}$ as follows:

\begin{itemize}
    \item Agreement: no two correct processes decide differently.
    
     For both $v \in \{0,1\}$, the following holds: ~$\forall k, k^\prime \in \mathbb{N}_0. ~$\\
     \begin{align}\label{formula:agree}
         \mathbf{A}(\mathbf{F}\bigvee_{\ell \in \mathcal{D}^n_v}\vb{\kappa}[\ell,k]>0 \rightarrow \mathbf{G}\bigwedge_{\ell^\prime \in \mathcal{D}^n_{1-v}}\vb{\kappa}[\ell^\prime, k^\prime]=0)  \tag{\textit{Agree}}
     \end{align}
     
    \item Validity: if all correct processes have $v$ as the initial value, then no process decides $1 - v$.       For both $v \in \{0,1\}$, the following holds: ~$\forall k \in \mathbb{N}_0. ~$\\
    \begin{align}\label{formula:valid}
        \mathbf{A}(\mathbf{G}\bigwedge_{\ell \in \mathcal{I}^n_v}\vb{\kappa}[\ell,0]=0 \rightarrow \mathbf{G}\bigwedge_{\ell^\prime \in \mathcal{D}^n_{v}}\vb{\kappa}[\ell^\prime, k]=0) \tag{\textit{Valid}}
    \end{align}

    \item Almost-sure Termination under Round-rigid Adversaries:  For every initial configuration $s$ and every round-rigid adversary $a$, the following holds:
    \begin{align}\label{formula:astermin}
        \mathbb{P}_{a}^{s} {[~\exists k \in \mathbb{N}_{0}.} 
        ~\mathbf{G} \bigwedge_{\ell \in \mathcal{F}^n \backslash \mathcal{D}^n} \kappa[\ell,k]=0~]=1 \tag{\textit{Termin}}
    \end{align}
\end{itemize}

\section{Verification of Randomized Consensus}
\label{sec:verification-consensus}

\subsection{Towards verifying non-probabilistic properties}\label{sec:nonpro-to-oneround}

For the specifications of safety properties, i.e., \emph{Agreement} and \emph{Validity}, we observe that they are both non-probabilistic properties and concern about locations in the threshold automata of correct processes. \emph{Agreement} contains two round variables $k$ and $k^\prime$, and \emph{Validity} considers round $0$ and $k$. We would like to check these specifications in the ByMC tool, which allows the properties to use only one round number. Therefore, we introduce two round invariants that refer to one round and prove that these two round invariants imply the consensus properties \emph{Agreement} and \emph{Validity} as follows.

The first round invariant claims that in every round and in every path, once a correct  process decides $v \in \{0,1\}$ in a round, no  correct process ever enters a location from $\mathcal{F}^n_{1-v}$ in that round. Formally, $\forall k \in \mathbb{N}_0. ~$

\begin{align}\label{inv:1}
    \mathbf{A} ( \mathbf{F} \bigvee_{\ell \in \mathcal{D}_v^n} \kappa[\ell, k]>0 ~\to~ \mathbf{G} \bigwedge_{\ell^\prime \in \mathcal{F}^n_{1-v} }\kappa[\ell^\prime,k] = 0 ) \tag{\textit{Inv1}}
\end{align}

The second round invariant claims that in every round and in every path, if no correct process starts a round with a value $v \in \{0,1\}$, then no correct process ever ends that round with $v$. Formally, $\forall k \in \mathbb{N}_0. ~$

\begin{align}\label{inv:2}
    \mathbf{A} ( \mathbf{G} \bigwedge_{\ell \in \mathcal{I}^n_v} \kappa[\ell, k]=0 ~\to~ \mathbf{G} \bigwedge_{\ell^\prime \in \mathcal{F}^n_{v} }\kappa[\ell^\prime,k] = 0 ) \tag{\textit{Inv2}}
\end{align}

Round switch lemma is useful in the following reasoning. It states that in every round and in every run,
if no process ever enters a final location with value $v$, then in
the next round, there will be no process in any initial location
with value $v$. 
\begin{lemma}[Round switch]
    For every \textit{Sys}$=\mathit{Sys}_\infty$(\textbf{TA}$^n$, \textbf{TA}$^c$) and every $v \in \{0,1\}$, we have: $~\forall k \in \mathbb{N}_0. ~$
    \begin{align}\label{lemma:roundswitch}
         \mathbf{A} ( \mathbf{G} \bigwedge_{\ell \in \mathcal{F}^n_v} \kappa[\ell, k] =0 ~\to~  \mathbf{G} \bigwedge_{\ell^\prime \in \mathcal{I}^n_v} \kappa[\ell^\prime, k+1] =0 ) \tag{\textit{RS}}
    \end{align}
\end{lemma}
\begin{proof}
    It follows from the definitions of $\mathcal{I}_v$, $\mathcal{F}_v$ and $\mathcal{B}_v$.
\end{proof}

\begin{proposition}\label{prop:agreevalid}
    If $\mathit{Sys} \models (\ref{inv:1}) \wedge (\ref{inv:2})$, then $\mathit{Sys} \models (\ref{formula:agree}) \wedge (\ref{formula:valid})$.
\end{proposition}

The proof is omitted due to length constraints.

We utilize a similar method as~\cite{VRCAURRA:BertrandKLW:21} to check one-round properties. The main idea is to prove that there exists a counterexample to the property in the multi-round system if and only if there is a counterexample in a single-round system. To this end, we prove the following two theorems in the non-probabilistic extended counter systems $\textit{Sys} _{\infty}  (\mathbf{TA}^{n},\mathbf{TA}^{c})$.

{
The first theorem
states that  every finite schedule can be reordered into a round-rigid one that is stutter equivalent regarding \textnormal{LTL}$_{-\textnormal{X}}$ formulas over proposition from $\textnormal{AP}_k$, for all rounds $k$, therefore it is sufficient to reason about the round-rigid schedules.}
\begin{theorem}\label{thm:roundrigid-schedule}

For every configuration $c$ and every finite
schedule $\tau$ applicable to $c$, there is a round-rigid schedule $\tau^\prime$ such that the following holds:
\begin{itemize}
    \item Schedule $\tau^\prime$  is applicable to configuration $c$,
    \item $\tau^\prime$ and $\tau$ reach the same configuration when applied to $c$, i.e., $\tau^\prime(c) = \tau(c)$,
    \item for every $k \in \mathbb{N}_0$, we have that $\textit{path}(c,\tau^\prime) $ and $  \textit{path}(c,\tau)$ are stutter equivalent w.r.t. $\textnormal{AP}_k$.
    % $\textit{path}(c,\tau^\prime) \triangleq_k  \textit{path}(c,\tau)$.
\end{itemize}
\end{theorem}

{The next theorem provides a simple way to check specifications with one round number, which include both round invariants \eqref{inv:1} and \eqref{inv:2}. 
It allows us to check specifications using single-round systems.}
{
First we need to build the single-round  threshold automata for $\mathbf{TA}^{n}$ and $\mathbf{TA}^{c}$.
\begin{definition}
    Given a $\mathbf{TA}$$=$$(\mathcal{L,V,R})$ over \textit{Env}, its single-round threshold automaton is $\mathbf{TA}_{rd}$$=$$(\mathcal{L}$$\cup$$ \mathcal{B}^\prime,\mathcal{V},\mathcal{R}_{rd})$, where $\mathcal{B}^\prime $$=$$ \{\ell^\prime$$: $ $\ell \in \mathcal{B}\}$ are copies of border locations, and the set of transition rules $\mathcal{R}_{rd}$$=$ $(\mathcal{R} $$\setminus$$ \mathcal{S}) \cup \mathcal{S}^\prime \cup \mathcal{R}_{loop}$, where $\mathcal{R}_{loop} $$=$ $ \{(\ell^\prime, \ell^\prime, \textit{true}, \vb{0})\}$ are self-loops at locations in $\mathcal{B}^\prime$, and $\mathcal{S}^\prime $$=$$ \{(\textit{from}, \ell^\prime, \textit{true}, \vb{0})$$: $ $(\textit{from}, \ell^\prime, \textit{true}, \vb{0}) \in \mathcal{S} $ with $\ell^\prime \in \mathcal{B}^\prime\}$ consists of modifications of round-switch rules.
\end{definition}
 }

{
Intuitively, given a multi-round $\mathbf{TA}$, we can remove its round-switch rules, add two more border locations $\mathcal{B}^\prime$ and rules that direct final locations to $\mathcal{B}^\prime$ and self-loops. In this way, we get a single-round threshold automaton $\mathbf{TA}_{rd}$ starting from  $\mathcal{B}$ and ending with $\mathcal{B}^\prime$, which models in some round a correct process starts and ends with a binary value, respectively. Given the single-round threshold automata for correct processes ($\mathbf{TA}^{n}_{rd}$) and common coin ($\mathbf{TA}^{c}_{rd}$), in a similar way, we can construct a single-round counter system $\mathit{Sys}^k\left(\mathbf{TA}^{n}_{rd},\mathbf{TA}^{c}_{rd}\right)$ to model the system in round $k$.
}

\begin{theorem}\label{thm:oneround}

Let $\mathrm{TA}$ be non-blocking, and let all fair executions of  $\textit{Sys}^{0}\left(\mathbf{TA}^{n}_{rd},\mathbf{TA}^{c}_{rd}\right)$   terminate w.r.t. all possible initial configurations. If  $\varphi[k]$ is a \textnormal{LTL}$_{-\textnormal{X}}$ formula over  $\mathrm{AP}_{k}$  for a round variable  $k \in \mathbb{N}_{0}$ , the following points are equivalent:
\begin{itemize}
    \item%[$(\mathrm{A})$] 
    $\textit{Sys} _{\infty}  (\mathbf{TA}^{n},\mathbf{TA}^{c})  \models \forall k \in \mathbb{N}_{0}. \mathbf{A} \varphi[k]$ 
    \item%[$(\mathrm{B})$]  
    $\textit{Sys}^{0}\left(\mathbf{TA}^{n}_{rd},\mathbf{TA}^{c}_{rd}\right) \models \mathbf{A} \varphi[0]$  with respect to the initial configurations $ \Sigma^{u}$, where $\Sigma^u$ is the union of all renamed initial configurations from all rounds.
\end{itemize}

\end{theorem}

The proofs are omitted due to length constraints. 
 % Theorem~\ref{thm:oneround} provides us a simple way to check specifications with one round number of the $\forall k \in \mathbb{N}_{0}. \mathbf{A} \varphi[k]$, which include the round invariants \eqref{inv:1} and \eqref{inv:2}. It allows us to check specifications using single-round systems, which can be checked automatically by ByMC tool.

\subsection{Round-rigid probabilistic termination}\label{subsec:termin}

Considering the differences in the design of common-coin-based consensus protocols,
we can roughly divide them into three categories. For each category of protocols, they have similar sufficient conditions for their almost-sure termination. 

\begin{figure}[tbp]
    \centering
    \includegraphics[width=0.25\textwidth]{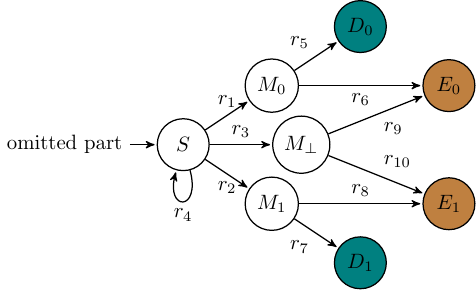}
    \caption{The Common Part in the Threshold Automata of (C)}
    \label{fig:ta-beforerefine}
\end{figure}

Specifically, let $\mathit{Sys}=\mathit{Sys}_\infty(\mathbf{TA}^n, \mathbf{TA}^c)$  and $\emph{Env}=(\Pi, RC, N)$ be the  counter system and environment of the protocol, and $\mathit{nc}$ is the number of modeled correct processes, we can classify the protocols in the following way:
\begin{enumerate}
    \item[$(A)$] the protocol does not have a ``decide'' action, that is, there are no accepting locations $\mathcal{D}$ in $\textbf{TA}^n$, the threshold automata of correct processes;
    \item[$(B)$] the protocol has a ``decide'' action and all its messages and conditions contain only binary values, that is, there are accepting locations $\mathcal{D}$ in $\textbf{TA}^n$, and for every pair of shared variables on the same messages $s_0,s_1 \in \Gamma$ and every final configuration $c_f \in \mathcal{F}$, $c_f.s_0 + c_f.s_1 \leq \mathit{nc}$;
    \item[$(C)$] the protocol has a ``decide'' action and uses a ``Binary Crusader Agreement'' primitive, that is, there are accepting locations $\mathcal{D}$ in $\textbf{TA}^n$, and there is a common part in $\textbf{TA}^n$ as shown in Fig.~\ref{fig:ta-beforerefine};
\end{enumerate}

% {\color{red} rephrase this para} Binary Crusader Agreement (BCA) [7] is a weak form of consensus, where processes start with a
% value in {0, 1} and can return a value in {0, 1, ⊥} (note the special value ⊥). The requirements
% are: (1) validity: if all correct processes start with the same input, then this is the only
% output, (2) agreement: no two correct processes output two distinct non-⊥ values, and
% (3) termination: every correct process eventually outputs a value. It is weaker than
% consensus because a process can output the “don’t know” value ⊥ instead of one of the inputs.
% Abraham et al.~\cite{EASABA:AbrahamBY:22} propose to leverage BCA as a primitive to construct a Byzantine consensus protocol, and we find that earlier protocols in~\cite{SFABC:MostefaouiMR:14,ABABug:miller:18} can be viewed as containing a BCA primitive.

\subsubsection{Sufficient conditions for $(A)$ protocols}

In this category of protocols, there is no ``decide'' action, and the \emph{Almost-sure Termination} property is stated as ``the probability of not all correct processes having the same value in $R$ round is $O(2^{-R})$''. The proof target of \emph{Almost-sure Termination} property for $(A)$ protocols can be formalized as follows.
For every initial configuration $c$ and every round-rigid adversary $a$, the following holds:
\begin{align}\label{formula:1Etermin}
        \mathbb{P}_{a}^{c} {[~\exists k \in \mathbb{N}_0, \exists v \in \{0,1\}.} 
        ~\mathbf{G} \bigwedge_{\ell \in \mathcal{F}^n_{1-v} } \kappa[\ell,k]=0~]=1  
\end{align}

We need two sufficient conditions to prove their \emph{Almost-sure Termination} property under round-rigid schedules:

\begin{itemize}
    \item  $(C1)$ states the existence of a positive probability lower bound for all processes ending round $k$ with the same final values. Formally, if there is a probability bound $p \in (0,1]$, such that for every round-rigid adversary $a$, every $k \in \mathbb{N}_0$, and every configuration $c_k$ that is initial for round $k$, it holds that
    \[ \mathbb{P}_{a}^{c_k} {[~\exists v \in \{0,1\}.} 
        ~\mathbf{G} \bigwedge_{\ell \in \mathcal{F}^n_{1-v} } \kappa[\ell,k]=0~] \geq p ~.  \]

    \item $(C2)$  states that if all correct processes start round $k$ with the same value $v \in \{0,1\}$, then they will all end with $v$ in that round. Formally, $\forall v \in \{0,1\}, \forall k \in \mathbb{N}_0. $
    \[   \mathbf{A} ( \mathbf{G} \bigwedge_{ \ell \in \mathcal{I}_{v}^n } \kappa[\ell,k]=0 ~\to ~ 
    \\
    \mathbf{G} \bigwedge_{\ell^\prime \in \mathcal{F}_{v}^n   } \kappa[\ell^\prime,k] = 0  ) ~. \]
\end{itemize}

Combining conditions $(C1)$ and $(C2)$, under every round-rigid adversary, from any initial configuration of round $k$, the probability
that all correct processes end with the same value in that round is at
least $p$,  and it holds for any future round. Thus, the probability not to have the same value within $n$ rounds is at most $(1-p)^n$, which tends to $0$ when $n$ tends to infinity.

\begin{proposition}
    If \textit{Sys} $\models (C1)$ and \textit{Sys} $\models (C2)$,then \textit{Sys} $\models \eqref{formula:1Etermin}$.
\end{proposition}
The proof is trivial. 

\subsubsection{Sufficient conditions for $(B)$ protocols} 

The protocols in this category can decide a binary value, and their messages contain only binary values. Their desired \emph{Almost-sure Termination} property requires that all correct processes decide the same value $v \in \{0,1\}$ with probability 1, and it can be stated as follows:

For every initial configuration $c$ and every round-rigid adversary $a$, the following holds:

\begin{align}\label{formula:1Dtermin}
        \mathbb{P}_{a}^{c} {[~\exists k \in \mathbb{N}_{0}, \exists v \in \{0,1\}.} 
        ~\mathbf{G} \bigwedge_{\ell \in \mathcal{F}^n \backslash \mathcal{D}^n_{v} } \kappa[\ell,k]=0~]=1  
\end{align}

Similarly we have to introduce three sufficient conditions to prove this specification:
\begin{itemize}
    \item $(C1)$ states the existence of a positive probability lower-bound for all processes ending round $k$ with the same final values. Formally, if there is a probability bound $p^\prime \in (0,1]$, such that for every round-rigid adversary $a$, every $k \in \mathbb{N}_0$, and every configuration $c_k$ that is initial for round $k$, it holds that
    \[ \mathbb{P}_{a}^{c_k} {[~\exists v \in \{0,1\}.} 
        ~\mathbf{G} \bigwedge_{\ell \in \mathcal{F}^n_{1-v} } \kappa[\ell,k]=0~] \geq p ~.  \]

    \item $(C2^\prime)$ states  the existence of a positive probability lower-bound for all correct processes deciding the same value $v \in \{0,1\}$ in round $k$
    if all correct processes start round $k$ with the same value $v$. Formally, if there is a probability bound $p^\prime \in (0,1]$, such that for every round-rigid adversary $a$, every $k \in \mathbb{N}_0$, and every configuration $c_k$ that is initial for round $k$,  it holds that: 
    
    $\forall v \in \{0,1\}, \forall k \in \mathbb{N}_0. $
    \[ \mathbb{P}_a^{c_k} [ \mathbf{G} \bigwedge_{ \ell \in \mathcal{I}_{1-v}^n } \kappa[\ell,k]=0 \to  \mathbf{G} \bigwedge_{\ell \in \mathcal{F}^n \backslash \mathcal{D}^n_{v}} \kappa[\ell^\prime,k] = 0 ] \geq p^\prime. \]

    % \item $(C3)$ coincides with round invariant (\ref{inv:2}), that is, if all correct processes start round $k$ with the same binary value $v$,  then they end with $v$ in round $k$. Formally, $\forall k \in \mathbb{N}_0. ~$

    % \begin{align*}
    %    \mathbf{A} ( \mathbf{G} \bigwedge_{\ell \in \mathcal{I}^n_v} \kappa[\ell, k]=0 ~\to~ \mathbf{G} \bigwedge_{\ell^\prime \in \mathcal{F}^n_{v} }\kappa[\ell^\prime,k] = 0 )
    % \end{align*}
\end{itemize}
{
Note that local-coin-based protocols with ``decide'' steps in ~\cite{VRCAURRA:BertrandKLW:21} share the same sufficient condition $(C2)$ for probabilistic termination with category $(A)$ protocols, while category $(B)$ protocols require a probabilistic condition $(C2^\prime)$. The difference lies in the guard of such ``decide'' steps: in a local-coin-based protocol, a correct process decides a binary value when it receives enough messages of certain type; however, in category $(B)$ (and also $(C)$) protocols, it additionally requires the value equals the common coin (see Line 10 in Fig.~\ref{fig:algo-mmr14}). }

\begin{proposition}
    Assume that there are accepting locations $\mathcal{D}$ in $\textbf{TA}^n$, and for every pair of shared variables on the same messages $s_0,s_1 \in \Gamma$ and every final configuration $c_f \in \mathcal{F}$, $c_f.s_0 + c_f.s_1 \leq \mathit{nc}$.
    If \textit{Sys} $\models (C1)$ and \textit{Sys} $\models (C2^\prime)$ 
    % and \textit{Sys} $\models (C3)$
    , then \textit{Sys} $\models (\ref{formula:1Dtermin})$.
\end{proposition}
\begin{proof}
    Let us fix the environment $\textit{Env}=(\Pi, RC, N)$, an initial configuration $c_0$ of \textit{Sys} and a round-rigid adversary $a$. 

    Two possible options may occur along a path $\pi \in \textit{paths}(c_0,a)$: 
    \begin{itemize}
        \item[$(a)$] $\exists v \in \{0,1\}. \pi \models \mathbf{G}( \bigwedge_{\ell \in \mathcal{F}^n_{v}} \kappa[\ell,0]= 0 )$,
        \item[$(b)$] $\forall v \in \{0,1\}.\pi \models \mathbf{F}( \bigvee_{\ell \in \mathcal{F}^n_{v}} \kappa[\ell,0] > 0 )$,
    \end{itemize}
    
    In words, either round $0$ ends with a final configuration where all correct processes have the same value, or round $0$ ends with a final configuration where both $0$ and $1$ present.
    In case $(a)$, we have $\pi \models \mathbf{G}( \bigwedge_{\ell \in \mathcal{F}^n_{v}} \kappa[\ell,0]= 0 )$ and by $(C1)$, for round $k=0$, the probability that this case happens is at least $p$. 
    % Then by Round-switch lemma \eqref{lemma:roundswitch}, we also have $\pi \models \mathbf{G}( \bigwedge_{ \ell^\prime \in \mathcal{I}^n_{v} } \kappa[\ell^\prime,1] = 0 )$. Using $(C3)$ and Round-switch lemma repeatly, we have that all correct processes propose the same value and end with the same value in any future round, that is, $\forall k \geq 1. \pi \models \mathbf{G}( \bigwedge_{\ell \in \mathcal{I}^n_{v}} \kappa[\ell,k]= 0 ) \bigwedge \mathbf{G}( \bigwedge_{\ell^\prime \in \mathcal{F}^n_{v}} \kappa[\ell^\prime,k]= 0 )$.

    % The probability that the second case occurs is at most $1-p$. In this case, round 1 starts with an initial configuration $c_1$ where the correct processes have both 0 and 1. From $c_1$ against adversary $a$, using the same reasoning as from $c_0$, at the end of round 1 we have two analogous cases and all correct processes end with the same value with probability at least $p$.

    % By iterating the reasoning, we have that with propability 1 all correct processes eventually end with the same binary value in some round. For $k \in \mathbb{N}_0$, consider the event $\mathcal{E}_k$: from $c_0$ under adversary $a$, not all correct processes end with the same value in the first $k$ round.  In particular, at the end of every round $i < k$ it is not the case that all correct processes have the same value. By the reasoning above, namely the second case for round $i$, this happens with probability at most $(1-p)$. Therefore, for $k$ rounds  we have  $\mathbb{P}^{c_0}_a[ \mathcal{E}_k ] \leq (1-p)^k$. % The limit is $0$ when $k$ tends to infinity.
    Consider the next round. By 
    % Assume that all correct processes end with the same value $v$ in round $k_0$ in the path $\pi_0$, that is, $\pi_0 \models \mathbf{G}( \bigwedge_{\ell \in \mathcal{F}^n_{1-v}} \kappa[\ell,k_0]= 0 )$. Applying $(C3)$ and 
    Round switch lemma \eqref{lemma:roundswitch}, we have that in round 1
    % all correct processes propose the same value $v$ and end with $v$ in any future round, that is, $\forall k \geq k_0+1. \pi_0 \models \mathbf{G}( \bigwedge_{\ell \in \mathcal{I}^n_{1-v}} \kappa[\ell,k]= 0 ) \bigwedge \mathbf{G}( \bigwedge_{\ell^\prime \in \mathcal{F}^n_{1-v}} \kappa[\ell^\prime,k]= 0 )$. 
    % Observe that in round $k_0 +1$, 
    all correct processes start with the same value $v$, two possible options may occur: 
    \begin{itemize}
        \item[$(c)$] $\pi_0 \models \mathbf{G} \bigwedge_{\ell \in \mathcal{F}^n \backslash \mathcal{D}^n_{v}} \kappa[\ell^\prime,1] = 0$,
        \item[$(d)$] $\pi_0 \models \mathbf{F} \bigvee_{\ell \in \mathcal{F}^n \backslash \mathcal{D}^n_{v}} \kappa[\ell^\prime,1] > 0$,
    \end{itemize}
    
    In words, either all correct processes decide $v$ in this round, or it ends with a final configuration where not all correct processes decide. According to $(C2^\prime)$, for round 1 the probability that case $(c)$ happens is at least $p^\prime$, if case $(a)$ happens in round 0. 

    Combine the first two rounds together. If case $(a)$ happens in round 0 and case $(c)$ happens in round 1, all correct processes decide the same binary value at the end of round 1, and its probability is at least $p \cdot p^\prime$. Therefore the probability that not all correct processes decide the same binary value at the end of first two rounds is at most $(1- p \cdot p^\prime)$.
    
    {
    By iterating the reasoning, consider the event $\mathcal{E}(2r)$: not all correct processes decide the same binary value at the end of first $2r$ rounds. Thus for round $2n$, we have that $$\mathbb{P}^{c_0}_a[\mathcal{E}(2n)] \leq (1- p \cdot p^\prime)^n$$
    \begin{equation}
        \lim_{n \to +\infty} \mathbb{P}^{c_0}_a[\mathcal{E}(2n)] = 0
    \end{equation}}
    The limit when $n$ tends
    to infinity yields that the probability of not having round-rigid termination is $0$. In conclusion, all correct processes decide the same binary value with probability 1.
\end{proof}

\subsubsection{Sufficient conditions for $(C)$ protocols} 
The \emph{Almost-sure Termination} property of $(C)$ protocols shares the same formula as \eqref{formula:1Dtermin}. However, they are based on \emph{Binary Crusader Agreement}, a weaker version of the binary consensus, 
{which introduces the third value $\bot$ for ``not sure about 0 or 1'', and the value $\bot$ is usually encoded by sending/receiving valid messages containing both values 0 and 1.}
%therefore the \emph{binding} property is a necessary condition for their termination.

Fig. \ref{fig:ta-beforerefine} shows the common structure of a single-round non-probabilistic threshold automaton for correct processes in the $(C)$ protocols.
Besides the final locations $\mathcal{F}=\{E_0,E_1,D_0,D_1\}$, there are 3 locations representing the output of \textit{Binary Crusader Agreement} 
primitive. We denote the set of locations $\{M_0, M_1, M_{\bot}\}$ by $\mathcal{M}$.
When receiving enough messages containing either a binary value or $\bot$ tagged with type $M$, it enters a location in $\mathcal{M}$  based on the number of messages received (encoded by variables $m_0, m_1$ and $m_{\bot}$). {There are two templates for the guard of $r_3$: $m_0 + m_1 \geq n-t-f$, and $m_0 + m_1 + m_{\bot} \geq n-t-f$.}
The transitions $r_5-r_{10}$ are all coin-based rules. $r_5,r_8, r_9$ share the same coin-based guard $cc_0 > 0$, which stands for the common coin result $0$, while $r_6,r_7, r_{10}$ share the same coin-based guard $cc_1 > 0$.

Here we introduce the property \emph{binding}, which is proposed in~\cite{EASABA:AbrahamBY:22}. Abstractly, a protocol obtains the binding property if no matter what the adversary does, it is forced to choose (bind to) in the present in a way that restricts all future outcomes of the protocol.
\begin{definition}[Binding]
   Let time $\tau$ be the first time such that there is a party that is correct and enters a location in $\mathcal{M}$ at time $\tau$. At time $\tau$, there is a value $b \in \{0,1\}$ such that no correct party enters $M_{1-b}$ in any extension of this execution.
   
   Formally, given a finite path $\pi$, we denote the set of all paths that take $\pi$ as its prefix by $\textit{Exts}(\pi)$. For every initial configuration $c_0$, every round-rigid adversary $a$ and every round $k$, the following holds:
    
   \[\forall \pi \in \textit{paths}(c_0,a).~ \pi \models \mathbf{F} \bigvee_{\ell \in \mathcal{M}} \kappa[\ell,k] >0 ~\to~ \]
   \[ ( \exists b \in \{0,1\}, \forall \pi^\prime \in \textit{Exts}(\pi). ~\pi^\prime \models \mathbf{G} \kappa[M_{1-b},k] = 0 )\]
\end{definition}

The \textit{binding} property states that for every execution prefix that some correct processes enter a location in $\mathcal{M}$ in round $k$, there is
a single binary value $b$ such that no correct process can enter location $M_{1-b}$  in this round in any future extension of this prefix. Note that this is an instance of a \emph{hyperproperty} because it characterizes sets of executions, i.e., all possible extensions of a prefix, instead of individual executions as in standard safety or liveness properties. % {\color{red} add the challenge of describing and verifying hyperproperty?}

It is natural to require that if a correct process enters $M_0$ in round $k$ then no correct process ever enters $M_1$, that is,  $\mathbf{A} (\mathbf{F} \kappa[M_0,k] $$>$$0 \to \mathbf{G} \kappa[M_1,k] $$=$$0 )$ (and similarly for $M_1$). However, it is different for the cases where a correct process enters $M_{\bot}$, because the transition $r_3$ from $S$ to  $M_{\bot}$ {concerns the total number of messages instead of the exact number of each message}, and the propositions on the numbers of messages are not supported in the threshold automata approach. Therefore, we propose a solution to further refine the model, particularly the transition rule to $M_{\bot}$, and encode the \emph{binding} property by five sufficient conditions. %{\color{red} or claim that this is a general solution/template for (C) protocols to refine the model and verify the conditions}

\begin{figure}[tbp]
    \centering
    \includegraphics[width=0.25\textwidth]{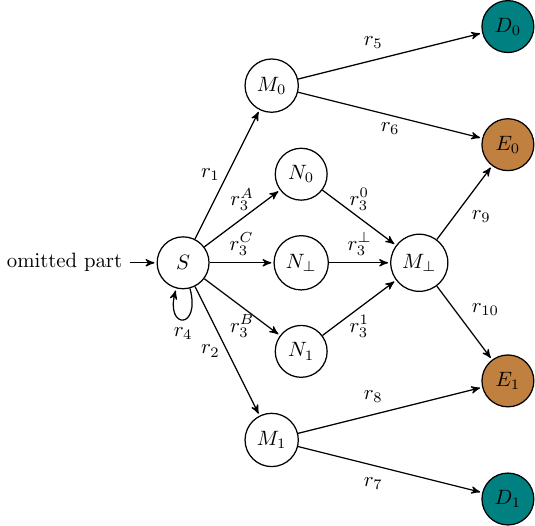}
    \caption{A Refined Model of the Common Part}
    \label{fig:ta-afterrefine}
\end{figure}

Assume that $r_3 = (S, M_{\bot}, \varphi, \vb{0})$, that is, it is a transition rule from location $S$
 to location $M_{\bot}$, guarded by formula $\varphi$, and it keeps the shared variables unchanged. We can remove $r_3$  and instead add three locations $N_0, N_1,$ and $N_{\bot}$ as well as the following transition rules:
 \begin{itemize}
     \item $r_3^A = (S, N_0, \varphi \wedge m_0 > 0, \vb{0})$
     \item $r_3^B = (S, N_1, \varphi \wedge m_1 > 0, \vb{0})$
     \item $r_3^C = (S, N_{\bot}, \varphi \wedge m_0 = 0 \wedge m_1 = 0, \vb{0})$
     \item $r_3^{i} = (N_{i}, M_{\bot}, \textit{true}, \vb{0})$, for $i \in \{0,1,\bot\}$
 \end{itemize}
 It is easy to observe that the modification does not block the automaton. By reasoning about the counter values of location $N_0$, $N_1$ and $N_\bot$, we can infer the number of messages received, which is impossible in the original automaton.  

Here we list the sufficient conditions for \textit{binding} property as follows:

\begin{itemize}
    \item \textit{(CB0)}: if a correct process enters $M_0$ in round $k$, then no correct process ever enters $M_1$, that is, $\forall k \in \mathbb{N}_0. ~$  $\mathbf{A} (\mathbf{F} \kappa[M_0,k] >0 ~\to~ \mathbf{G} \kappa[M_1,k] =0 )$;
    \item \textit{(CB1)}: if a correct process enters $M_1$ in round $k$, then no correct process ever enters $M_0$, that is, $\forall k \in \mathbb{N}_0. ~$ $\mathbf{A} (\mathbf{F} \kappa[M_1,k] >0 ~\to~ \mathbf{G} \kappa[M_0,k] =0 )$;
    \item \textit{(CB2)}: if a correct process enters $N_0$ and then $M_{\bot}$  in round $k$, then no correct process ever enters $M_1$, that is, $\forall k \in \mathbb{N}_0. ~$ $\mathbf{A} (\mathbf{F} \kappa[N_0,k] >0 ~\to~ \mathbf{G} \kappa[M_1,k] =0 )$;
    \item \textit{(CB3)}: if a correct process enters $N_1$ and then $M_{\bot}$  in round $k$, then no correct process ever enters $M_0$, that is, $\forall k \in \mathbb{N}_0. ~$ $\mathbf{A} (\mathbf{F} \kappa[N_1,k] >0 ~\to~ \mathbf{G} \kappa[M_0,k] =0 )$;
    \item \textit{(CB4)}: if a correct process enters $N_{\bot}$ and then $M_{\bot}$  in round $k$, then no correct process ever enters $M_0$ or $M_1$, that is, $\forall k \in \mathbb{N}_0. $ $\mathbf{A} (\mathbf{F} \kappa[N_{\bot},k] >0 \to \mathbf{G} \bigwedge_{\ell \in \{ M_0, M_1 \}} \kappa[\ell,k] =0 )$;
\end{itemize}

\begin{proposition}\label{prop:binding}
    If $\textit{TA}^n$ has such a subpart as Fig.~\ref{fig:ta-afterrefine} and \textit{Sys} satisfies all (CB0)-(CB4), then \textit{Sys} satisfies Binding property.
\end{proposition}
\begin{proof}
    Let time $\tau$ be the first time such that there is a party that is correct and enters a location in $\mathcal{M}$ at time $\tau$.
    There are five possible options for the correct process:
    \begin{itemize}
        \item it enters $M_0$ through the transition rule $r_1$, then by (CB0) we can set the binary value $b=0$ and no correct process can enter location $M_{1-b} = M_1$, that is, $\mathbf{G} \kappa[M_1]=0$;
        \item it enters $M_1$ through the transition rule $r_2$, then by (CB1) we can set the binary value $b=1$ and no correct process can enter location $M_{1-b} = M_0$, that is, $\mathbf{G} \kappa[M_0]=0$;
        \item it enters $M_{\bot}$ through the transition rule $r_3^A$, then by (CB2) we can set the binary value $b=0$ and no correct process can enter location $M_{1-b} = M_1$;
        \item it enters $M_{\bot}$ through the transition rule $r_3^B$, then by (CB3) we can set the binary value $b=1$ and no correct process can enter location $M_{1-b} = M_0$;
        \item it enters $M_{\bot}$ through the transition rule $r_3^C$, then by (CB4) we can set the binary value $b=0$ and no correct process can enter location $M_{1-b} = M_1$;
    \end{itemize}
    To conclude, at time $\tau$, there must be a value $b \in \{0,1\}$ such that no correct party enters $M_{1-b}$ in any extension of this execution.
\end{proof}

\begin{proposition}\label{prop:c-termin}
    Assume that there are accepting locations $\mathcal{D}$ in $\textbf{TA}^n$, and there is a subpart in $\textbf{TA}^n$ as shown in Fig.~\ref{fig:ta-afterrefine}.
    If $\textit{Sys}$ satisfies Binding property as well as $(C2^\prime)$,  %and $(C3)$, 
     then $\textit{Sys}  \models (\ref{formula:1Dtermin})$.
\end{proposition}

\begin{proof}
    Fix the environment $\textit{Env}=(\Pi, RC, N)$, a round-rigid adversary $a$ and an initial configuration $c_0$. {Assume that the common coin is $\epsilon$-\text{Good}, that is, for any value $v \in \{0,1\}$, the common coin result equals $v$ with probability $\geq \epsilon >0$.} 

% $\epsilon$-\emph{Good} is an important property of the common coin abstraction: for any value $v \in \{0,1\}$, all
% correct parties output $v$ with probability $\geq \epsilon$. If a coin is $\frac{1}{2}$-good, we call it a \emph{strong coin}. In this paper, we consider only the protocols that employ strong coins.

% In round 0, there are three possible cases: the first correct process that enters a location in $\mathcal{M}$ actually enters $M_0$, $M_1$ or $M_{\bot}$. 
{
Consider any round $k$, let time $\tau_k$ be the first time such that there is a process that is correct and enters a location in $\mathcal{M}$ at time $\tau_k$. 
By \textit{binding} property, at time $\tau_k$ we can have a binary value $b$ such that no correct process can ever enter $M_{1-b}$ in this round. Because all coin-based rules $(r_5 - r_{10})$ start from a location in $\mathcal{M}$, no correct process has ever executed a coin-based rule at time $\tau_k$. In other words, at time $\tau_k$ the common coin of round $k$ is not yet tossed, therefore the common coin result is independent of the binary value $b$.
Note that the common coin result has at least $\epsilon$ possibility of being value $b$.
If no correct process can ever enter $M_{1-b}$ in  round $k$ and the common coin result is $b$, all correct processes will have the same value $b$ at the end of round $k$. To conclude, by \textit{binding} property, there is a positive probability lower bound $\epsilon$ for all correct process ending round $k$ with the same final value, and it coincides the condition $(C1).$ } 

Follow the same aforementioned reasoning, and we prove that with probability 1 all correct processes decide the same binary value.
\end{proof}
% Assume that all correct processes have the same estimate value $v$ at the end of round $k_0$ in the path $\pi_0$, that is, $\pi_0 \models \mathbf{G}( \bigwedge_{\ell \in \mathcal{F}^n_{1-v}} \kappa[\ell,k_0]= 0 )$. Using $(C3)$ and Round-switch lemma \eqref{lemma:roundswitch}, we have that all correct processes propose the same value $v$ and end with $v$ in any future round.
% Applying $(C2^\prime)$ we see that two possible options may occur in round $k_0+1$: either all correct processes decide $v$ in this round, or it ends with a final configuration where some do not decide. Following the same reasoning, we have that with probability 1 all correct processes decide the same value. 

% Combine the two phases together and follow the same calculation, we have that even in the worst cases that all correct
% processes must first reach a configuration of ending with the same value and then decide the same value, all correct processes decide with probability 1. 

\begin{corollary}
    Assume that there are accepting locations $\mathcal{D}$ in $\textbf{TA}^n$, and there is a subpart in $\textbf{TA}^n$ as shown in Fig.~\ref{fig:ta-beforerefine}.
    If \textit{Sys} satisfies all (CB0)-(CB4) conditions, as well as $(C2^\prime)$ %and $(C3)$
    , then $\textit{Sys} \models \eqref{formula:1Dtermin}$.
\end{corollary}

\begin{proof}
    It follows directly by Propositions~\ref{prop:binding} and~\ref{prop:c-termin}.
\end{proof}

\subsubsection{Reducing probabilistic to non-probabilistic
specifications}

Note that the sufficient conditions $(C2)$, %$(C3)$, 
as well as \textit{(CB0)}-\textit{(CB4)}, are non-probabilistic specifications with one round number, so that we can check them using the method in Sect. \ref{sec:nonpro-to-oneround}; while $(C1)$ and $(C2^\prime)$ are probabilistic, we need to further reduce its verification to verification of non-probabilistic specifications.

{The method for local coins in~\cite{VRCAURRA:BertrandKLW:21} requires that there exists at most one ``coin-based'' rule in a threshold automaton and it must lie at the end of a round. We remove those restrictions on the ``common-coin-based'' rules. In fact, it is even allowed that a single-round path in a threshold automaton contains multiple ``common-coin-based'' rules. This makes the extended model more expressive, while it keeps good properties, such as the following lemma. }%The following lemma.}

% Two sufficient conditions of \emph{Almost-sure Termination} properties $(C1)$ and $(C2^\prime)$ are probabilistic and we cannot check them directly in the ByMC tool. We need the following lemma to check non-probabilistic properties instead.

\begin{lemma}\label{lemma:bound}
 Let \textit{Env} be the environment, $\mathcal{A}^R$ be the set of round-rigid adversaries and $\mathcal{I}$ be the set of initial configurations over \emph{Env}. In the single-round  probabilistic counter system $\textit{Sys}(\mathbf{TA}^{n}_{rd},\mathbf{TA}^{c}_{rd})$, for every \textnormal{LTL}$_{-\textnormal{X}}$ formula $\varphi$ over atomic proposition
AP, the following statements are equivalent:
\begin{itemize}
    \item%[(A)] 
    $\exists p >0, \forall c \in \mathcal{I}, \forall a \in \mathcal{A}^R.  ~\mathbb{P}^c_a(\varphi) \geq p$,
    \item%[(B)] 
    $\forall c \in \mathcal{I}, \forall a \in \mathcal{A}^R, \exists \pi \in {\textit{paths}(c,a)}. ~\pi \models \varphi$
\end{itemize}
\end{lemma}

The proof is omitted due to length constraints.

By applying Lemma~\ref{lemma:bound}, we observe that the probabilistic sufficient conditions are equivalent to non-probabilistic specifications with the existential quantifier in the single round system $\textit{Sys}(\mathbf{TA}^{n}_{rd},\mathbf{TA}^{c}_{rd})$. We can further turn them into non-probabilistic specifications that can be checked by ByMC, and in the end prove the sufficient conditions $(C1)$ and $(C2^\prime)$.

\begin{table*}[tbp]
    \centering
    \caption{Benchmarks of 8 Different Common Coin-Based Protocols}
    % \begin{tabular}{ c c c c c c c c c c c}
    %      \hline
    %      \multicolumn{5}{c}{Automaton}  & \multicolumn{2}{c}{inv1} & \multicolumn{2}{c}{inv2} & \multicolumn{2}{c}{a.s. termin}  \\ \hline
    %      % \cline{1-4} %\cline{4-5} \cline{6-7} \cline{8-9}
    %      Name & category & $|\mathcal{L}|$ & $|\mathcal{R}|$ & $|\mathcal{M}|$ & nschemas & time  & nschemas & time  & nschemas & time (total) \\ \hline
    %      \texttt{Rabin83} & $(A)$& 7 & 17 & 0 & 6 & 0.25  & 2 & 0.20  & 8 & 0.43 \\ \hline
    %      \texttt{CC85(a)} & $(B)$& 9 & 18 & 3 & 342 & 4.93  & 42 & 0.50 & 171.5 & 3.36 \\
    %      \texttt{CC85(b)} & $(B)$ & 10 & 17 & 0 & 6 & 0.25  & 2 & 0.20  &  8 & 0.40 \\
    %      % \texttt{cachin}00 & 1+D & 9 & 17  & 31.0  & 3.7  & 23 \\
    %      \texttt{FMR05} &$(B)$& 10 & 16 & 0 & 6 & 0.23 & 2 &  0.21 & 2 & 0.41 \\ 
    %      \texttt{KS16} & $(B)$ & 11 & 26 & 1 & 18 & 0.75 & 5 & 0.31  & 15 & 1.02 \\
    %      \hline
    %      \texttt{MMR14} & $(C)$ & 17 & 29 & 6 & 28918 & 298.90 & 1442  & 8.74 & - &  CE \\
    %      \texttt{Miller18}(mpi) & $(C)$ & 22 & 48 & 10 & $> 10^6$ & 605  & 253534 & 226 & $> 10^8$ & 16h08m18s \\ 
    %      \texttt{ABY22}(mpi) & $(C)$ & 22 & 49 & 10 & $> 10^6$ & 583  & 106098 & 71 & $> 10^8$ & 13h36m57s \\ 
    %      \hline
         
    % \end{tabular} 
    \begin{tabular}{ c c c c c c c c c c}
         \hline\hline
         \multicolumn{4}{c}{Automaton}  & \multicolumn{2}{c}{Agreement} & \multicolumn{2}{c}{Validity} & \multicolumn{2}{c}{A.S. Termination}  \\ \hline
         % \cline{1-4} %\cline{4-5} \cline{6-7} \cline{8-9}
         Name & category & $|\mathcal{L}|$ & $|\mathcal{R}|$  & nschemas & time  & nschemas & time  & nschemas & time (total) \\ \hline
         \texttt{Rabin83} & $(A)$& 7 & 17  & 6 & 0.25  & 2 & 0.20  & 8 & 0.43 \\ \hline
         \texttt{CC85(a)} & $(B)$& 9 & 18  & 342 & 4.93  & 42 & 0.50 & 171.5 & {2.70} \\
         \texttt{CC85(b)} & $(B)$ & 10 & 17  & 6 & 0.25  & 2 & 0.20  &  8 & {0.32} \\
         % \texttt{cachin}00 & 1+D & 9 & 17  & 31.0  & 3.7  & 23 \\
         \texttt{FMR05} &$(B)$& 10 & 16  & 6 & 0.23 & 2 &  0.21 & 2 & {0.32} \\ 
         \texttt{KS16} & $(B)$ & 11 & 26  & 18 & 0.75 & 5 & 0.31  & 15 & {0.76} \\
         \hline
         \texttt{MMR14} & $(C)$ & 17 & 29  & 28918 & 298.90 & 1442  & 8.74 & - &  CE \\
         \texttt{Miller18}(mpi) & $(C)$ & 22 & 48 & $> 10^6$ & 605  & 253534 & 226 & $> 10^8$ & {11h46m47s} \\ 
         \texttt{ABY22}(mpi) & $(C)$ & 22 & 49  & $> 10^6$ & 583  & 106098 & 71 & $> 10^8$ & {10h13m14s} \\ 
         \hline
         
    \end{tabular} 
    
    \label{tab:benchmark-prop}
\end{table*}

\begin{table}[tbp]
\centering
\caption{Properties Checked for Value $0$}
% \begin{tabular}{c c c}
% % \begin{tabular}{ *{1}{>{\raggedright\arraybackslash}m{0.1\textwidth}} *{1}{>{\raggedright\arraybackslash}m{0.25\textwidth}} *{1}{>{\raggedright\arraybackslash}m{0.2\textwidth}} *{1}{>{\raggedright\arraybackslash}m{0.5\textwidth}}}
% \hline \hline
%   Label& Name  & Formula\\
%   \hline
%  \eqref{inv:1} & inv1  & $\mathbf{A}~\mathbf{F}(\textsc{Ex}\{D_0\}) ~\rightarrow~ \mathbf{G}~ (\neg \textsc{Ex}\{E_1, D_1\} ) $\\
%  % \textbf{A1} & agree\_1 & N & $\mathbf{A}~\mathbf{F} ~\textsc{Ex}\{E_1, D_1\} ~\rightarrow~ \mathbf{G}~ (\neg \textsc{Ex}\{E_0, D_0\} ) $ \\
%  \eqref{inv:2} & inv2  & $\mathbf{A}~ \textsc{All}\{I_0\} ~\rightarrow~ \mathbf{G}~( \neg \textsc{Ex}\{E_1, D_1\} )  $\\ 
%  % \textbf{V1} & valid\_1 & N & $\mathbf{A}~ \textsc{All}\{I_1\} ~\rightarrow ~\mathbf{G}( \neg \textsc{Ex}\{E_0, E_1, D_0\} )  $\\
%  % \textit{(RT)} & round\_term  & $\mathbf{A}~ \textit{fair} ~\rightarrow~ \mathbf{F}~ \textsc{All} \{D_0, D_1, E_0, E_1\}$\\
%  %\textbf{RT}$^\prime$ & round-term$^\prime$  & $\mathbf{A}~ \textit{fair} ~\rightarrow~ \mathbf{F}~ \textsc{All} \{M_0, M_1, M_{\bot}\}$\\
%  $(C1)$ &  termin\_1 & $\mathbf{A}~ \mathbf{F}( \textsc{Ex}\{D_0, E_0\}) ~\to~ \mathbf{G}(\neg \textsc{Ex}\{D_1, E_1\})$ \\
%  \textit{(CB0)} & binding\_0  & $\mathbf{A}~ \mathbf{F}( \textsc{Ex}\{M_0\}) ~\to~ \mathbf{G}( \neg \textsc{Ex}\{M_1\} )$ \\
%  %\textbf{C3} &   &  $\mathbf{A}~ \textsc{All}\{I_0\} ~\rightarrow~ \mathbf{G}~( \neg \textsc{Ex}\{M_1, M_{\bot}\} )  $ \\
%  $\cdots$ & & \\
% \hline
% \end{tabular}
\begin{tabular}{c c}
\hline \hline
  Label&  Formula\\
  \hline
 \eqref{inv:1} &  $\mathbf{A}~\mathbf{F}(\textsc{Ex}\{D_0\}) ~\rightarrow~ \mathbf{G}~ (\neg \textsc{Ex}\{E_1, D_1\} ) $\\
 \eqref{inv:2} &  $\mathbf{A}~ \textsc{All}\{I_0\} ~\rightarrow~ \mathbf{G}~( \neg \textsc{Ex}\{E_1, D_1\} )  $\\ 
 $(C1)$ &   $\mathbf{A}~ \mathbf{F}( \textsc{Ex}\{D_0, E_0\}) ~\to~ \mathbf{G}(\neg \textsc{Ex}\{D_1, E_1\})$ \\
 \textit{(CB0)}  & $\mathbf{A}~ \mathbf{F}( \textsc{Ex}\{M_0\}) ~\to~ \mathbf{G}( \neg \textsc{Ex}\{M_1\} )$ \\
 % \textit{(C3)}  & $\mathbf{A}~ \mathbf{G}( \neg \textsc{Ex}\{I_0\}) ~\to~ \mathbf{G}( \neg \textsc{Ex}\{M_0\} )$ \\
 $\cdots$ & $\cdots$ \\
\hline
\end{tabular}
\label{tab:expri-properties}
\end{table}

\section{Experiments}
\label{sec:experiments}

We have applied our approach to eight randomized fault-tolerant consensus protocols that make use of common coins, including:

\begin{enumerate}
    % \item Randomized consensus in~\cite{AVFC:Ben-Or:83} with clean crashes (\texttt{ben-or-cc}), for which a process either sends to all or none. This protocol uses local coins and works correctly when $n>2t$.
    % \item Randomized Byzantine consensus in~\cite{AVFC:Ben-Or:83} (\texttt{ben-or-byz}). This protocol uses local coins and tolerates $t$ Byzantine faults when $n>5t$.
    \item Randomized Byzantine consensus (\texttt{Rabin83}) in~\cite{RBG:Rabin:83}, the first common coin-based randomized consensus protocol. It tolerates Byzantine faults when $t<n/10$.
    % \item Randomized consensus in~\cite{ABAP:Bracha:87} (\texttt{bracha}). It uses local coins and runs as a high-level protocol together with a reliable broadcast primitive that reduces the impact of Byzantine faults to ''little more than fail-stop faults''. We check only the high-level protocol for clean crashes with resilience condition $n>3t$.
    \item Randomized Byzantine consensus (\texttt{CC85(a)}) in~\cite{SERBAA:ChorC:85}, which proposes a simple implementation of common coin and has an optimal resilience condition $n>3t$.
    \item Randomized Byzantine consensus (\texttt{CC85(b)}) in~\cite{SERBAA:ChorC:85}, which is an adaptation of \texttt{Rabin83} and raises the bound of Byzantine faults to $t<n/6$.
    % \item Randomized Byzantine agreement in~\cite{ROCPABA:CachinKS:05} (\texttt{cachin}). This protocol makes use of threshold signatures and cryptographic common coin-tossing protocols, which makes it difficult to model its decision game.  It tolerates up to $t<n/3$ Byzantine faults
    \item Randomized Byzantine agreement (\texttt{FMR05}) in~\cite{SEORCP:FriedmanMR:05}, which contains one communication step in each round, and can resist up to $n/5 $ Byzantine faults.
    \item Randomized Byzantine agreement (\texttt{KS16}) in~\cite{BAEPT:King:16}, which builds on Bracha's~\cite{ABAP:Bracha:87} and replaces the local coin in each process with a common coin. Its resilience condition remains $n>3t$.
    \item The protocol (\texttt{MMR14}) by Most\'efaoui et al.~\cite{SFABC:MostefaouiMR:14}, which contains an attack by adaptive adversary resulting in non-termination.
    \item In Miller's post~\cite{ABABug:miller:18} there is a discussion to fix the bug of  \texttt{MMR14}, and the fixed version (\texttt{Miller18}) was later used in the Dumbo protocol~\cite{DBLP:conf/ccs/GuoL0XZ20}.
    \item Randomized Byzantine agreement (\texttt{ABY22}) in~\cite{EASABA:AbrahamBY:22} based on binding crusader agreement, which tolerates $t$ Byzantine faults when $n>3t$.

    %\item  Randomized Byzantine agreement in~\cite{EASABA:AbrahamBY:22} (\texttt{abraham-gbca}) which weakens the assumption of the common coin and makes use of graded binding crusader agreement~\cite{OPPSBA:FeldmanM:97} instead. It tolerates $t$ Byzantine faults  when $n>3t$.
\end{enumerate}

% Following the reduction approach of \ref{sec:reducetoslgames}, we encode two versions of one-round non-probabilistic threshold automata for each benchmark except \texttt{cachin}: one for verifying the winning strategy of its proposal game, the other for verifying the winning strategy of its decision game. Both automata are provided as the input to Byzantine Model Checker (ByMC)~\cite{BMC:KonnovW:18}. 

% \ref{tab:expri-properties} gives a summary of the verified properties in our benchmarks. gGiven the set of all possible locations $\mathcal{L}$ and a subset $S=\{ \ell_1, \ldots, \ell_n  \} ~\subseteq~ \mathcal{L}$, we adopt the shorthand notation for LTL$_{-\mathrm{x}}$ formulae:  $\textsc{Ex}\{ \ell_1, \ldots, \ell_n \}$ stands for $\bigvee_{ \ell \in S } \kappa[\ell] \neq 0$, that is, at least one location in $S$ has a non-zero counter value ; $\textsc{All}\{\ell_1, \ldots, \ell_n\}$ stands for $\bigwedge_{ \ell \in \mathcal{L}\backslash S } \kappa[\ell]=0$, that is, any location out of $S$ has a counter value of $0$.
% Though only the properties for value $0$ are listed, we also need to check for value $1$, because \textit{luck}'s winning strategies may not be symmetric for binary values.

%\renewcommand{\arraystretch}{1.25}

For each protocol, we build two versions of one-round threshold automata for probabilistic and non-probabilistic properties. We input both automata into ByMC, which implements the parameterized model checking techniques. 

Table~\ref{tab:expri-properties} gives a summary of the properties and conditions that were verified in our experiments. The round invariants inv1 and inv2 are sufficient conditions for \emph{Agreement} and \emph{Validity}, and the others are sufficient conditions for \emph{Almost-sure termination}. 
% \textcolor{modified}{We adopt the shorthand notation for LTL$_{-\mathrm{x}}$ formulae
% }
Given the set of all possible locations $\mathcal{L}$ and a subset $S=\{ \ell_1, \ldots, \ell_n  \}$, we adopt the shorthand notation for \textnormal{LTL}$_{-\textnormal{X}}$ formulae: $\textsc{Ex}\{ S \}$ stands for $\bigvee_{ \ell \in S } \kappa[\ell] \neq 0$, that is, at least one threshold automaton enters a location in $S$; %at least one location in $S$ has a non-zero counter value; 
$\textsc{All}\{S\}$ stands for $\bigwedge_{ \ell \in \mathcal{L}\backslash S } \kappa[\ell]=0$, that is, all threshold automata enter the locations in $S$. %any location out of $S$ has a counter value of $0$. 
Take \eqref{inv:2} as example, its formula states that for every execution where all correct processes propose value $0$ in a round, it is always true that no process can end with value $1$ in that round.

Table~\ref{tab:benchmark-prop} shows the computational results of the experiments: column $|\mathcal{L}|$ and $|\mathcal{R}|$ give the numbers of automata locations and rules. In the columns for properties,  ``nschemas'' stands for the number of checked schemas of executions. The computational times are given in seconds if not specified, and ``CE'' stands for a reported counterexample. The benchmarks of category (C) are challenging for the technique of Konnov et al.~\cite{ASCPTA:KonnovLVW:17}: the sizes of their threshold automata are too large, and as a result their experiments timeout for a 24-hour limit.
Therefore we check the first six rows with a laptop with Intel Core i7-12650H, while the last two experiments are conducted on a computing server of 216 cores AMD EPYC 7702 in parallel mode.  We ran each experiments 10 times and listed the average values for nschemas and time.

As can be seen in Table~\ref{tab:benchmark-prop}, the benchmarks of categories (A) and (B) have small sizes and they can be quickly verified for the three properties. 
A violation of \emph{Binding} sufficient conditions is found for \texttt{MMR14} within 10 seconds at best.% and 30 minutes at worst, and we chose not to list the average values due to their significant variability. 
The found counterexample contains the system settings, e.g. $n=193$ and $t=64$, an initial configuration as well as a sequence of actions. It reports that the resulted state does not satisfy the formula of  \textit{(CB2)}. We analyze its execution and find that it follows the pattern of the designed attack in~\cite{ABABug:miller:18}: firstly some correct processes access the common coin, then the adaptive adversary obtains the coin value $v$ and manipulates the messages, and finally a correct process gets its new \emph{est} value as $1-v$. By repeating the last actions we can complete the duality in the attack.
%, which can be seen as a concrete attack by an adaptive round-rigid adversary. 
\texttt{Miller18} and \texttt{ABY22}, two fixed versions of \texttt{MMR14}, both pass the check, while it takes over 11 and 10 hours to check all their sufficient conditions for \emph{Almost-sure Termination} properties. %\textcolor{added}{The longest checking of a single formula is approximately 4.5 hours in parallel mode.}

Table~\ref{tab:benchmark-prop} reveals a positive correlation between the number of checked schemas and the verification time, and the worst-case schemas depend on the size of the automaton and the structure of the formula. Given that a threshold automaton is tightly coupled with its protocol, reducing the number of locations and transition rules is not straightforward without sacrificing essential meaning. Another critical factor affecting the maximum number of schemas is the presence of  milestones. Intuitively, when a milestone is reached, certain threshold guards (inequations) are always true/false in the future executions. Interestingly, different types of messages lead to varying numbers of milestones. As an illustrative example, we modify \texttt{ABY22}  to create 5 threshold automata of the same size but with different milestone counts. %\footnote{Note that it is of only theoretical interest because the modified automata are not coupled with existing protocols.} 
The maximum numbers of schemas are then calculated and presented in Table~\ref{tab:scales}.

\begin{table}[tbp]
    \caption{Maximum Numbers of Schemas for Threshold Automata with Different Milestones}
    \centering
    \begin{tabular}{c | c  c c }
\hline \hline
      Name & Formula & nmilestones & max-nschemas \\
      \hline
       \texttt{ABY22}  & \textit{(CB0)} &  10 & 98182294 \\
       \texttt{ABY22}-1  & \textit{(CB0)} &  9 & 15129955 \\
       \texttt{ABY22}-2  & \textit{(CB0)} &  8 & 2650445 \\
       \texttt{ABY22}-3  & \textit{(CB0)} &  7 & 257126 \\
       \texttt{ABY22}-4  & \textit{(CB0)} &  6 & 28918 \\
       \hline
       \texttt{ABY22}  & \eqref{inv:2} &  10 & 7479057 \\
       \texttt{ABY22}-1  & \eqref{inv:2} &  9 & 1298630 \\
       \texttt{ABY22}-2  & \eqref{inv:2} &  8 & 253534 \\
       \texttt{ABY22}-3  & \eqref{inv:2} &  7 & 28395 \\
       \texttt{ABY22}-4  & \eqref{inv:2} &  6 & 3592 \\
       \hline
    \end{tabular}
    \label{tab:scales}
\end{table}
\section{Related Work}
\label{sec:relatedwork}

In recent years, there have been steady advance in creating ever more complex distributed consensus protocols, in order to optimize for performance while preserving its guarantees of fault-tolerance. HoneyBadger~\cite{DBLP:conf/ccs/MillerXCSS16} is the first practical asynchronous atomic broadcast protocol. The Dumbo family of protocols~\cite{DBLP:conf/ccs/GuoL0XZ20,DBLP:conf/ccs/Gao0L00Z22} build upon HoneyBadger to obtain further improvements in performance and latency. All these protocols use asynchronous Byzantine agreement as a substep. The increasing complexity of these protocols make it more urgent to develop scalable formal techniques for their verification.

There have been a long line of work on verification of fault-tolerant distributed protocols. Earlier work~\cite{AVRDCP:KwiatkowskaNS:01,VRBA:KwiatkowskaN:02} make use of Cadence SMV and the probabilistic model checker PRISM to verify protocols consisting of 10-20 processes. For parameterized verification, the key development is the proposal of threshold automata, and theoretical results reducing the correctness in the general case to model checking on a finite system~\cite{ASCPTA:KonnovLVW:17,DBLP:journals/iandc/KonnovVW17}. These results are implemented in the ByMC tool~\cite{BMC:KonnovW:18} for verifying protocols in practice.

Bertrand et al.~\cite{VRCAURRA:BertrandKLW:21} extend the framework of threshold automata with probabilistic transitions, namely probabilistic threshold automata, and reduce the safety and liveness verification of a multi-round randomized protocol to checking on a one-round automaton. In that paper, round-rigid adversaries are assumed in the proof of termination. This restriction is relaxed to that of weak adversaries in~\cite{DBLP:conf/vmcai/BertrandLW21}.

% \begin{enumerate}
%     \item probabilistic threshold automata restrict that the probabilistic transition rules only happen at the end of each round,  therefore the protocols that do not end with coin-tossing such as \texttt{cachin} cannot be modeled. We do not have such restriction on our model, and the probabilistic transitions are replaced by deterministic winning strategies;
    
%     \item probabilistic threshold automata work only for randomized protocols with local coins such as \texttt{ben-or-byz} and \texttt{bracha}, which have an exponential expected number of rounds and thus limited practical use, while our approach can be applied on both local and common coin-based protocols, and 6 out of 9 benchmarks in~\ref{sec:experiments} use common coins;
    
%     % \item Our approach provides further quantitative analysis of (worst case) expected round complexity as long as the winning strategies are verified.
% \end{enumerate}

In the direction of compositional reasoning, the work by Bertrand et al.~\cite{HVBC:BertrandGKL:22} verified the consensus protocol of Red Belly Blockchain by reducing the protocol into two parts: an inner broadcast protocol and an outer decision protocol, verified each part separately and then composed the correctness results. % \textcolor{added}{It could be used to reduce the overall verification time, but there is no general approach to reducing a complex protocol and proving the correctness of the composition. }

Besides the use of threshold automata, there is also work using Ivy and TLA proof system to verify a simplified version of HotStuff~\cite{DBLP:conf/forte/Jehl21}, a protocol for repeated distributed consensus used in permissioned blockchains~\cite{DBLP:conf/podc/YinMRGA19}. The work by Attiya et al.~\cite{FSRBPDAG:AttiyaEN:23} focused on verifying Byzantine fault-tolerant protocols based on directed acyclic graphs (DAGS), some of which also make use of randomness in the form of common coins.

\section{Conclusion}
\label{sec:conclusion}

In this paper, we proposed an extension of probabilistic threshold automata that supports the use of common coins, that can be used to model fault-tolerant randomized distributed protocols for reaching consensus. We reduce the correctness properties of consensus protocols: agreement, validity, and termination to queries on single-round non-probabilistic threshold automaton, which can be checked using ByMC. Key parts of the queries correspond to the binding condition preventing an adaptive adversary from delaying decision by the protocol indefinitely. Using this framework, we verified eight consensus protocols that make use of common coins, and were able to reproduce an attack found in earlier work.

Our work can be viewed as a further step toward verifying complex Byzantine consensus protocols% proposed in recent work
, such as HoneyBadger~\cite{DBLP:conf/ccs/MillerXCSS16} and Dumbo~\cite{DBLP:conf/ccs/GuoL0XZ20}, both of which make use of asynchronous Byzantine agreement. Future work includes the development of composition reasoning techniques as well as reasoning about cryptographic primitives that are part of those protocols. Finally, we intend to remove the restriction of round-rigid adversaries in the proof of termination, perhaps by using methods similar to those proposed in~\cite{DBLP:conf/vmcai/BertrandLW21}.

\section*{Acknowledgment}
% {
% We are grateful to Andrea Turrini for his constructive comments.
% We thank anonymous reviewers for their insightful feedback. 
This work is supported by CAS Project for Young Scientists in Basic Research, Grant No.YSBR-040, ISCAS New Cultivation Project  ISCAS-PYFX-202201, and ISCAS Basic Research ISCAS-JCZD-202302.
% \newline\protect\includegraphics[height=8pt]{EU.jpg}

This work is part of the European Union’s Horizon 2020 research and innovation programme under the Marie Sk\l{}odowska-Curie grant no.\@ 101008233.
% }

\bibliographystyle{IEEEtran}
\bibliography{ref}

\end{document}